\documentclass[preprintnumbers,superscriptaddress,nofootinbib,aps,prd,floatfix]{revtex4}

\pdfoutput=1
\usepackage{graphicx}
\usepackage{latexsym}
\usepackage{amsfonts}
\usepackage{amssymb}
\usepackage{amsmath}
\usepackage{slashed}
\usepackage{feynmp}
\usepackage{hyperref}
\usepackage{url}
\usepackage{color}
\usepackage{cancel}
\usepackage{multirow,array}
\usepackage{float}
\usepackage{subfigure}
  
\begin{document}
\title{Applying Liouville's Theorem to Gaia Data}

\author{Matthew R.~Buckley}
\affiliation{Department of Physics and Astronomy, Rutgers University, Piscataway, NJ 08854, USA}

\author{David W.~Hogg}
\affiliation{Center for Cosmology and Particle Physics, Department of Physics, New York University, 726 Broadway, New York, NY 10003, USA}
\affiliation{Center for Data Science, New York University, 60 Fifth Ave, New York, NY 10011, USA}
\affiliation{Max-Planck-Institut f\"{u}r Astronomie, K\"{o}nigstuhl 17, D-69117 Heidelberg, Germany}
\affiliation{Flatiron Institute, 162 Fifth Ave, New York, NY 10010, USA}

\author{Adrian M.~Price-Whelan}
\affiliation{Department of Astrophysical Sciences, Princeton University, 4 Ivy Lane, Princeton, NJ 08544, USA}
\affiliation{Flatiron Institute, 162 Fifth Ave, New York, NY 10010, USA}

\begin{abstract}
The Milky Way is filled with the tidally-disrupted remnants of globular clusters and dwarf galaxies. Determining the properties of these objects -- in particular, initial masses and density profiles -- is relevant to both astronomy and dark matter physics. However, most direct measures of mass cannot be applied to tidal debris, as the systems of interest are no longer in equilibrium. Since phase-space density is conserved during adiabatic phase mixing, Liouville's theorem provides a connection between stellar kinematics as measured by observatories such as {\it Gaia} and the original mass of the disrupted system. Accurately recovering the phase-space density is complicated by uncertainties resulting from measurement errors and orbital integration, which both effectively inject entropy into the system, preferentially decreasing the measured density. In this paper, we demonstrate that these two issues can be overcome. First, we measure the phase-space density of the globular cluster M4 in {\it Gaia} data, and use Liouville's theorem to derive its mass. We then show that, for tidally disrupted systems, the orbital parameters and thus phase-space density can be inferred by minimizing the phase-space entropy of cold stellar streams. This work is therefore a proof of principle that true phase-space density can be measured and the original properties of the star cluster reconstructed in systems of astrophysical interest.
\end{abstract}

\maketitle

\section{Introduction}

The Milky Way was assembled from hierarchical mergers of smaller, gravitationally-bound objects, themselves composed of stars, gas, and dark matter \cite{1965ApJ...142.1317P,1974ApJ...187..425P,1976MNRAS.177..717W,1978ApJ...225..357S,1984Natur.311..517B}.
Most of these building blocks were or are being tidally disrupted by the Galaxy, taking objects that were initially compact in position-space and distributing their mass throughout the Milky Way.
Depending on the age of a merger, the mass of the progenitor system, and the Galactic orbit of the object, this tidally-stripped material may still remain relatively compact in the form of ``stellar streams'' or other coherent phase-space structures \citep[e.g.,][]{1994Natur.370..194I, Lisanti:2011as, Johnston:1997fv,1998ApJ...495..297J, Johnston:2012yh, Malhan:2018, Price-Whelan:2018}, or may fill a substantial (positional) volume of the Galaxy \citep[e.g.,][]{1999MNRAS.307..495H, Myeong:2018}.
Thanks to large photometric, astrometric, and spectroscopic stellar surveys such as (and amongst others) the Sloan Digital Sky Survey \citep[SDSS;][]{York:2000, Eisenstein:2011, Blanton:2017} including SEGUE \citep{Yanny:2009} and APOGEE \citep{Majewski:2017}, the Dark Energy Survey \citep[DES;][]{DES:2016}, and the \textit{Gaia} mission \citep{2018A&A...616A...1G}, many such systems have been identified using kinematic and chemical information of Milky Way stars, enabling a detailed reconstruction of the formation history of our Galaxy \citep[e.g.,][]{Belokurov:2006, 2016ASSL..420.....N, Belokurov:2018, Helmi:2018}.

While these recent revelations about the assembly of the Milky Way are driving a new era of precision ``Galactic archaeology,'' the building blocks themselves -- former dwarf galaxies and globular clusters -- are interesting in their own right.
For one, the initial dark matter masses of these systems provides information about the clustering of matter on small scales, which is sensitive to dark matter particle physics \citep[e.g.,][]{Moore:1999, Hargis:2014, Kim:2018, Robles:2019}.
In more detail, the internal stellar dynamics of Milky Way satellite systems depends strongly on the distribution of dark matter in these substructures and should therefore provide stronger constraints on the physics and nature of dark matter \citep[e.g.,][]{Buckley:2017ijx}.
Current constraints on dark matter physics from surviving Milky Way satellites are somewhat inconclusive, in large part because bound satellites tend to be distant (in order to survive to present day) and thus present a significant challenge for obtaining precise 3D kinematics for individual stars in these systems. 
However, tidally-disrupted systems exist and are numerous throughout the inner Galaxy:
These ideas motivate the development of a technique to measure the initial properties of tidally-stripped objects {\it prior} to infall and merger with the Milky Way using only post-merger kinematic information.
This is a difficult task: tidal debris and substructures are no longer gravitationally bound and are therefore dispersed in position-space. Thus, most techniques for mass measurement (e.g., rotation curves or the virial theorem) are not applicable. 
However, the coherence in phase-space suggests a new method, which we will develop in this paper.

Most mass estimators of gravitationally bound systems rely on conserved quantities (for example, total energy or angular momentum) and assume some equilibrium has been achieved. When a star cluster is disrupted, the conservation laws can only be applied to the combined object -- both the disrupting system and the disrupter -- and the assumption that the system has reached equilibrium is generally not true. If we are to measure the mass of these systems, we need some new conserved quantity that can be related to the total mass even when the system is out of equilibrium. Liouville's theorem provides such a quantity. 

In adiabatic processes, Liouville's theorem tells us that the phase-space density $f$ is conserved:
\begin{equation}
\frac{d f(\vec{w})}{dt} = 0,
\end{equation}
where $\vec{w}$ is a six-dimensional set of canonical coordinates and momenta. For mergers of smaller halos that are less than ${\cal O}(0.1)$ of the Milky Way mass, the mergers are adiabatic \cite{BoylanKolchin:2003sf}, and so we expect Liouville's theorem to apply to the coarse-grained phase-space density of the merging component.

As we will show in this paper, given knowledge of the phase-space density of a system has evolved adiabatically from a state of gravitational equilibrium one can reconstruct structural properties of the original configuration, including the mass. Most crudely, the total phase-space volume of a gravitationally bound relaxed system (defined as the six-dimensional volume where $f(\vec{w})$ has support) is related to the system's total mass. This should make intuitive sense: all else being equal, when in gravitational equilibrium, a more massive system will have constituent particles extending out to larger radii and moving faster at a given radius than a less massive system, resulting in a larger phase-space volume. This will persist even after disruption, thanks to Liouville's theorem. Quantitatively, for a gravitationally-bound system, the six-dimensional phase-space volume $V_6$ is proportional to the total mass $M$ of a system as 
\[
V_6 \propto M^{3/2}.
\]
However, replacing the proportionality with an equality requires some assumptions about the parametrics of the overall density distribution when the system was gravitationally bound -- for example, we must know the concentration and scale length of the original system. In some cases, these additional parameters may be estimated by comparison with other similar systems of known mass, at the cost of systematic error. 

A more powerful approach is to consider the phase-space density of a set of tracer stars in the system, $\{ f_i\}$, rather than just the total phase-space volume extracted from this set. Under the assumptions that lead to Liouville's theorem, this set of density measurements is identical to the original density distribution $f(\vec{w},\vec{\xi})$, which depends on the system's structural parameters $\vec{\xi} = \{\xi_i\}$, specifying the profile shape (e.g., NFW or King), mass, scale lengths, {\it etc.} For a given $\vec{\xi}$, we can calculate the probability  $P(f_i|\vec{\xi})$ of measuring each $f_i$ given that they are being drawn from a distribution set by parameters $\vec{\xi}$. The true value of $\vec{\xi}$ can then be obtained by minimizing the negative log-likelihood,
\[
\lambda(\vec{\xi}) = -2\sum_i \ln P_i(f_i|\vec{\xi}).
\]
This allows for a measurement of the total mass without requiring prior knowledge of the other parameters of the density distribution included in $\vec{\xi}$.

Using these techniques, the original mass of a gravitationally-bound system being tidally disrupted in an adiabatic merger with a larger galaxy can be directly determined by measuring the phase-space density of a set of tracer stars (in any set of canonical coordinates) after the merger. Importantly, we do not need a complete sample of stars from the merged structure to effect this measurement: as long as there are a sufficient number of identified tracer stars which sample the probability density in an unbiased manner, then we can in principle recover the original density distribution and thus the mass. Phase-space density has been used to measure masses of non-disrupted structures, for example by using the caustic curve defined by the line-of-sight projection of the density distribution of galaxy clusters \cite{1997ApJ...481..633D,1999MNRAS.309..610D,2012MNRAS.426.2832A}. However, to our knowledge, this application of Liouville's theorem to directly measure the mass of astrophysical systems has not been considered previously. \\

Though Liouville's theorem leads to a conceptually simple connection between the present-day kinematics of a system and that of the pre-merger progenitor, the inevitable realities of measurement errors and uncertainty in the stellar orbits act as injections of entropy into the system. If not corrected for, these errors will completely swamp the phase space measurement, making this application of Liouville's theorem to achieve accurate  mass-measurement impossible. 

The entropy is related to the phase-space density $f$ by
\begin{equation}
S = - \int d^6 \vec{w} f(\vec{w}) \ln f(\vec{w}). \label{eq:entropy_def}
\end{equation}
Measurement uncertainties of the tracer star kinematics or using an incorrect model of the gravitational potential in which the system is orbiting will act to effectively increase the entropy of a given system as compared to reality. This will, on average, {\it decrease} the phase-space density and so {\it increase} the phase space volume and the inferred mass. 
We demonstrate that both of these effects can be accounted for, leading to a new dynamical method for measure initial masses of bound or previously bound stellar systems.

This paper serves as a proof-of-principle in applying Liouville's theorem to Milky Way stars using {\it Gaia} kinematic data with the ultimate goal of accurately measuring the mass of disrupted Galactic substructure. We demonstrate that two of the major sources of entropy in the measurement: kinematic errors and incorrectness of the Galactic potential used to compute orbits, can be corrected for, allowing the density distribution to be reconstructed and the mass determined. In Section~\ref{sec:volume}, we describe in detail how to related the mass of a system that was once in gravitational equilibrium to the phase-space densities of tracer stars. In Section~\ref{sec:mismeasure}, we consider the effect of realistic measurement errors, by considering the non-disrupted M4 globular cluster in the {\it Gaia} data. We show that, by considering the location of minima of the phase space volume as a function of the phase-space coordinates for each star, we can correct for these errors and make an accurate measurement of the mass. In Section~\ref{sec:orbits}, we consider the increase in entropy introduced by numerical integration of orbits, which is a necessary step in applying this technique to disrupted systems.

\section{Phase Space Volume and Mass \label{sec:volume}}

Liouville's theorem states that, under adiabatic evolution, the phase-space distribution function or density function, $f(\vec{w})$, acts as an incompressible conserved fluid.\footnote{The adiabatic requirement is made clear when one considers the definition of entropy $S$ in Eq.~\eqref{eq:entropy_def}.} Here, $\vec{w}$ is a six-vector of canonical coordinates $\vec{q}$ and conjugate momenta $\vec{p}$, e.g., position and velocity $\vec{w} = \{\vec{x},\vec{v}\}$. For disrupted systems orbiting in the Milky Way, we will see (in Section~\ref{sec:orbits}) that a useful alternative set of coordinates are the action-space integrals and their canonical-conjugate phase angles $\vec{w} = \{\vec{J},\vec{\theta}\}$. In any set of canonical coordinates, $f$ is conserved under adiabatic evolution.

Given sufficiently accurate measurements of the phase space coordinates $\vec{w}_i$ of a set of tracer stars from a gravitationally bound system, it is possible -- in principle -- to measure the phase-space density of each tracer. Deriving this density is of course a computationally difficult problem, especially in six-dimensions; a problem compounded by measurement errors. 

We will address these issues in Sections~\ref{sec:mismeasure} and \ref{sec:orbits}. In this Section, we investigate what we can learn about the properties of the original density distribution, if we assume that the phase-space density of individual tracer stars is known without errors. That is, we assume there is a known set of phase-space densities $\{ f_i \}$ (or phase-space number densities $\{ n_i \}$) for stars which were drawn from an (unknown) initial density distribution $f$.

As a concrete example, consider some small object (a dwarf galaxy or globular cluster) being accreted and tidally stripped by the Milky Way. In some approximation, the object begins as a self-gravitating bound configuration of mass, which evolves into an unbound object experiencing primarily the gravity of the Milky Way. During this process, the object moves from spherically symmetric and compact in position-space, and ends as a  ``stream'' dispersed in through the Galaxy \citep[e.g.,][]{Johnston:1998bd, 1999MNRAS.307..495H}. While the position-space volume increases, the velocity-space volume decreases, as all components of the object end up moving nearly in the same direction within the cold stream. Thus, the total six-dimension volume remains constant, assuming that the mass internal to the orbits of each star does not change significantly during a dynamical time \cite{1986ApJ...301...27B,2006ApJ...645..240P}.

There should therefore be a connection between the values of the phase-space density associated with each tracer star and the mass distribution of the progenitor system (i.e., when they were gravitationally bound). For example, the more massive at halo is, the faster stars can move at a given radius and remain bound. Alternatively, for fixed speed, a star can orbit at larger radii. Thus, the volume of phase space over which the density has support must increase as the halo mass increases, all other parameters being equal.

In order to make the connection between the phase-space distribution and the mass explicit, we first consider how the phase-space volume changes with mass (though we will eventually see that the full phase-space distribution has far more information that makes for a more powerful analytic tool).  Defining the phase-space volume as the integral over phase space where the phase-space density has support,
\begin{equation}
V_6 = \int d^6\vec{w} \,\Theta\left[f(\vec{w})\right],
\end{equation}
where $\Theta$ is the Heaviside theta function, we see that Liouville's theorem implies that $V_6$ is conserved. 

Using the physical coordinates $\vec{x}$ and $\vec{v}$, we first calculate the volume before any disruption has occurred for a spherically symmetric system where $f$ is non-zero only within a radius $r_{\rm max}$. A maximal estimate of the phase-space volume $V_6$ is then
\begin{equation}
V_6 = (4\pi)^2 \int_0^{r_{\rm max}} r^2 dr \int_0^{v_{\rm max}(r)} v^2 dv = 
\frac{32 \sqrt{2} \pi^2 G^{3/2} M^{3/2}}{3} \int_0^{r_{\rm max}} dr \,r^2 \left[ \int_r^{r_{\rm max}} \frac{\delta_M(r')}{(r')^2} dr' \right]^{3/2}. \label{eq:v6mass}
\end{equation}
Here, $v_{\rm max}(r)$ is the maximum velocity possible at a radius $r$ and $\delta_M(r)$ is the fraction of the total mass enclosed in radius $r$. 

Eq.~\eqref{eq:v6mass} makes manifest the $M^{3/2}$ dependence for the six-dimensional phase space volume. But as can be seen, the volume depends on a double integral over the density distribution. It is here that the particular functional form of the density distribution for the non-disrupted system is encoded. Additional information about the bound system is necessary to define this integral: For example, the scale radius $r_0$ and the concentration parameter $c$ for some spherical mass models. We will call the set of parameters which define the original density distribution $\vec{\xi}$, including the mass. Ideally, we would be able to derive the mass while scanning over the other parameters, but we see in Eq.~\eqref{eq:v6mass} that the total phase space volume only provides a measurement for the mass assuming the other elements of $\vec{\xi}$ are known.

That the other parameters must be known somewhat limits the utility of the total phase space volume to the problem of mass measurement, restricting it to situations where the structure of the system of interest when it was in gravitational equilibrium are already determined. This is possible if the system is not disrupted, or if the parameters can be estimated by comparison with known systems expected to have similar density profiles. However, in the former case, there exist other mass measurement methods that are simpler and more straightforward, given that the object is still gravitationally self-bound. In the latter, comparison with other systems introduces new systematic errors into the mass derivation.  \\ 

However, there is far more information in the set of phase-space density values than was used to construct $V_6$. After all, it is not just the total volume that is conserved under adiabatic evolution, but the individual densities. For example, if we imagine varying the concentration or scale radius of a stellar cluster, not only would the total phase space volume change, but so would the maximum phase-space density that can be obtained by any tracer star. One could then use the maximum measured density as an additional handle to constrain the parameters $\vec{\xi}$ of the system.

This suggests a way to utilize all of the information in the conserved phase-space density. Rather than considering only an integral or moment of the phase-space density, we consider the probability of measuring the set of phase-space densities themselves, $\{ f_i \}$, {\it given} an assumed $\vec{\xi}$. We can then recover the mass, concentration, scale radius, {\it etc} by optimizing the likelihood of observing $\{ f_i \}$ given these parameters. That is, we minimize the negative log-likelihood
\begin{equation}
- 2 \ln \lambda(\vec{\xi}) = -2 \sum_i \ln P(f_i|\vec{\xi}), 
\end{equation}
where the minimization is over the parameters in $\vec{\xi}$. 

Let us demonstrate the idea using a specific example: the King profile which has been demonstrated to be a good fit to globular clusters, quasi-spherical collections of stars with little or no dark matter \cite[e.g.,][]{Conroy:2010bs}. A King profile is defined by three parameters: a tidal radius $r_t$ beyond which there are no stars (i.e., $r_t \equiv r_{\rm max}$), a characteristic radius $r_0$ (the ``King radius''), and a characteristic speed $\sigma$. The speed can be rewritten as a function of the total mass $M$ and $r_0$, and the tidal radius as a concentration parameter $c \equiv \log_{10} r_t/r_0$. We will take $\vec{\xi} = \{M, r_0, c\}$ as our three free parameters for a King profile.

The King profile is given by
\begin{equation}
f_K(\vec{x},\vec{v}) = \left\{\begin{array}{cr} \frac{\rho_1}{(2\pi\sigma^2)^{3/2}} \left[ e^{E/\sigma^2} -1 \right] & E > 0 \\ 0 & E \leq 0 \end{array}\right. \label{eq:king}
\end{equation}
where the energy per mass is
\begin{equation}
E(r) = \Phi(r_t) - \Phi(r) - \frac{1}{2} v(r)^2,
\end{equation}
and the parameters $\sigma$ and $\rho_1$ can be expressed in terms of $\{ M, r_0, c\}$. 
The reduced potential 
\begin{equation}
W(r/r_0) = [\Phi(r_t/r_0) - \Phi(r/r_0)]/\sigma^2
\end{equation} 
can be analytically solved for in terms of the boundary condition $W(r_t/r_0) = 0$, and so the functional form of $W$ depends only on the concentration parameter $c$.

If a tracer star drawn from a King profile is measured to have a phase-space density $f_i \pm \delta f$, where $\delta f$ is some statistical or systematic error on the measurement, then this implies that, in the bound globular cluster, the star was in the velocity range
\begin{eqnarray}
v & \in & [ v_{\rm min},v_{\rm min}] \\
v_{\rm min}(r) & = & 2\sigma \left( W(r) - \ln\left[ \frac{(2\pi \sigma)^{3/2}}{\rho_1}(f_i+\delta f) +1 \right] \right)^{1/2} \nonumber \\
 v_{\rm max}(r) & = & 2\sigma \left( W(r) - \ln\left[ \frac{(2\pi \sigma)^{3/2}}{\rho_1}(f_i-\delta f) +1 \right] \right)^{1/2}. \nonumber
\end{eqnarray}
The probability of measuring a tracer star with this range of phase-space densities is then the integral of the density over this constrained range of velocities divided by the integral of the density over all positions and velocities (i.e., the total mass $M$):
\begin{equation}
P(f_i \pm \delta f|M,r_0,c) = M^{-1} \int_0^{r_t} 4\pi r^2 dr \int_{v_{\rm min}(r)}^{v_{\rm max}(r)}4\pi v^2 dv f_K(\vec{r},\vec{v}).
\end{equation}
This results in a numerically integrable equation for the probability of measuring the tracer star with a given phase-space density. Minimizing the log-likelihood can recover the true values of the total mass $M$, as well as the scale radius $r_0$, and concentration $c$. That is, we can profile over $\vec{\xi} = \{ M,r_0,c\}$ to minimize
\begin{equation}
-2 \ln \lambda(M,r_0,c) = -2 \sum_i \ln P(f_{i, \rm low},f_{i,\rm high} |M,r_0,c),
\end{equation}
where the probability $P$ for the $i^{\rm th}$ star is the probability that a star with phase-space density measured to be in the interval $[f_{i,\rm low},f_{i,\rm high}]$ is drawn from a distribution set by $\{M,r_0,c\}$.

Though the King profile allows for a clean factorization in terms of the velocity and position integrals, the probability can also be estimated by explicitly constructing the distribution of $f$ for a given set $\vec{\xi} = \{M,r_0,c\}$ using a large number of tracers drawn from a simulated profile, and then calculating the probability of finding a given $[f_{i,\rm low},f_{i,\rm high}]$ within that distribution. Computationally, we find the latter method is somewhat faster, though it also introduces errors due to finite sampling effects. This numerical technique also can be more easily generalized to density distributions where the constraint on the phase-space variables given a measurement of density is not so easily calculated as in the King profile.

\begin{figure}[t]
\includegraphics[width=0.36\columnwidth]{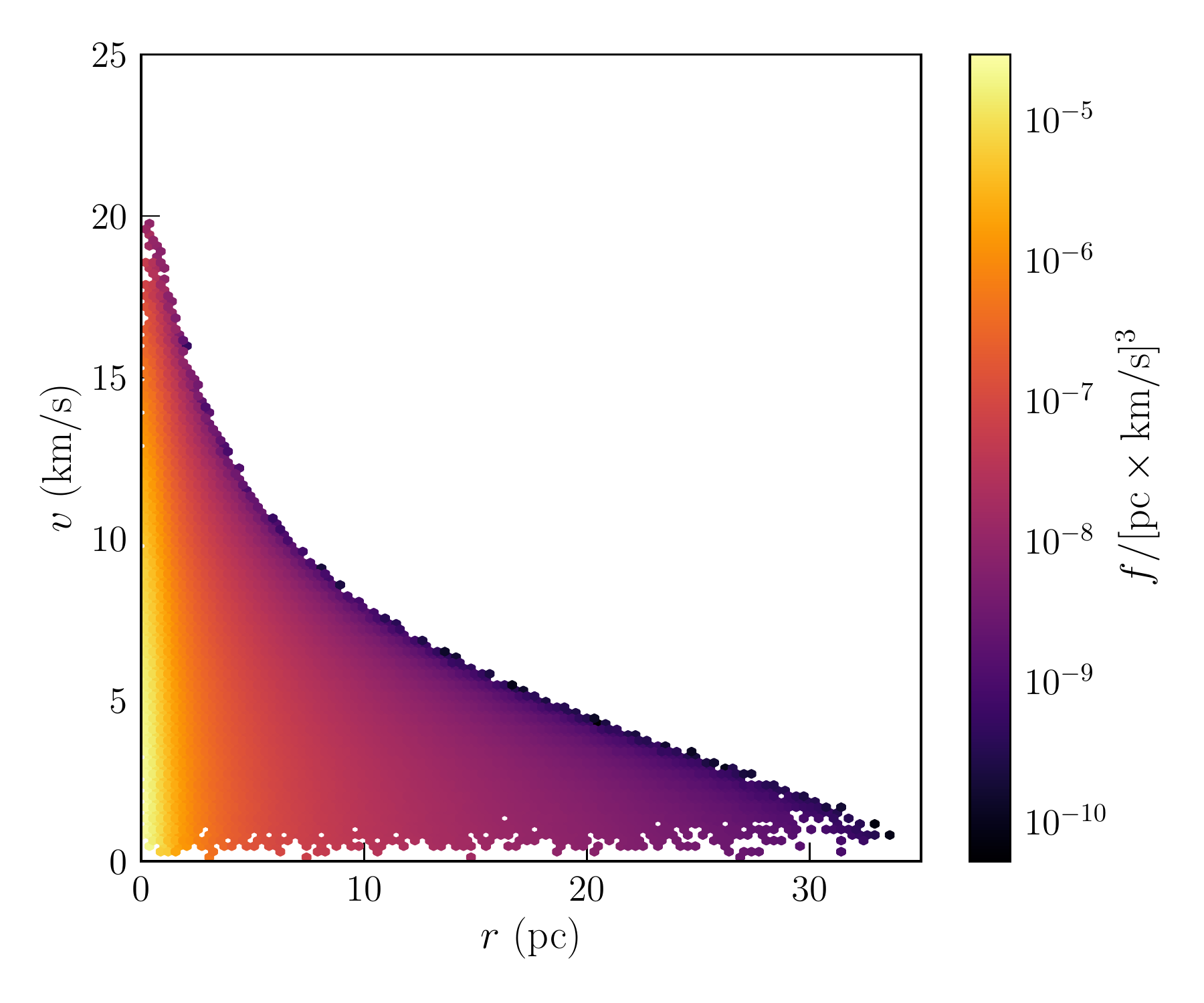}\includegraphics[width=0.3\columnwidth]{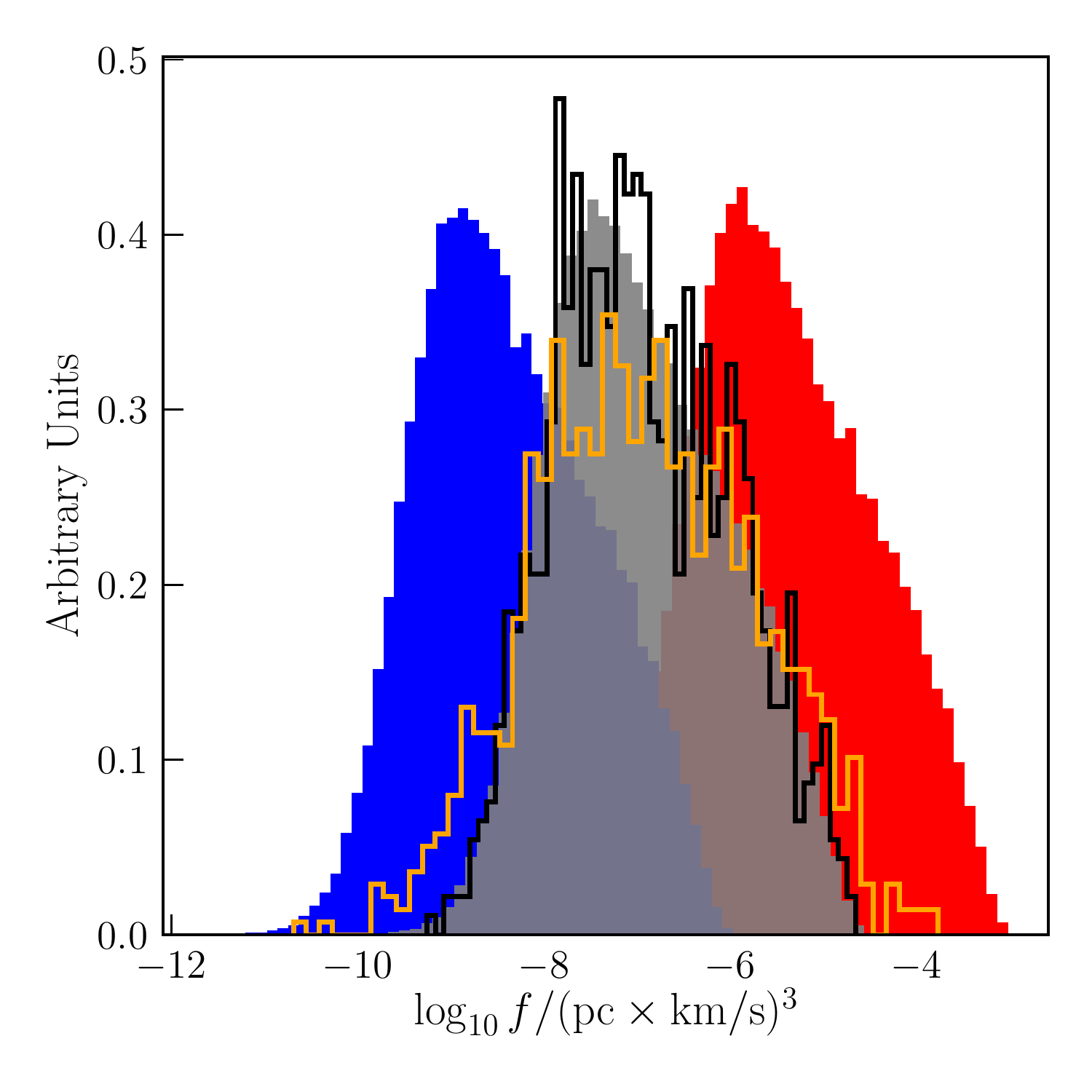}\includegraphics[width=0.3\columnwidth]{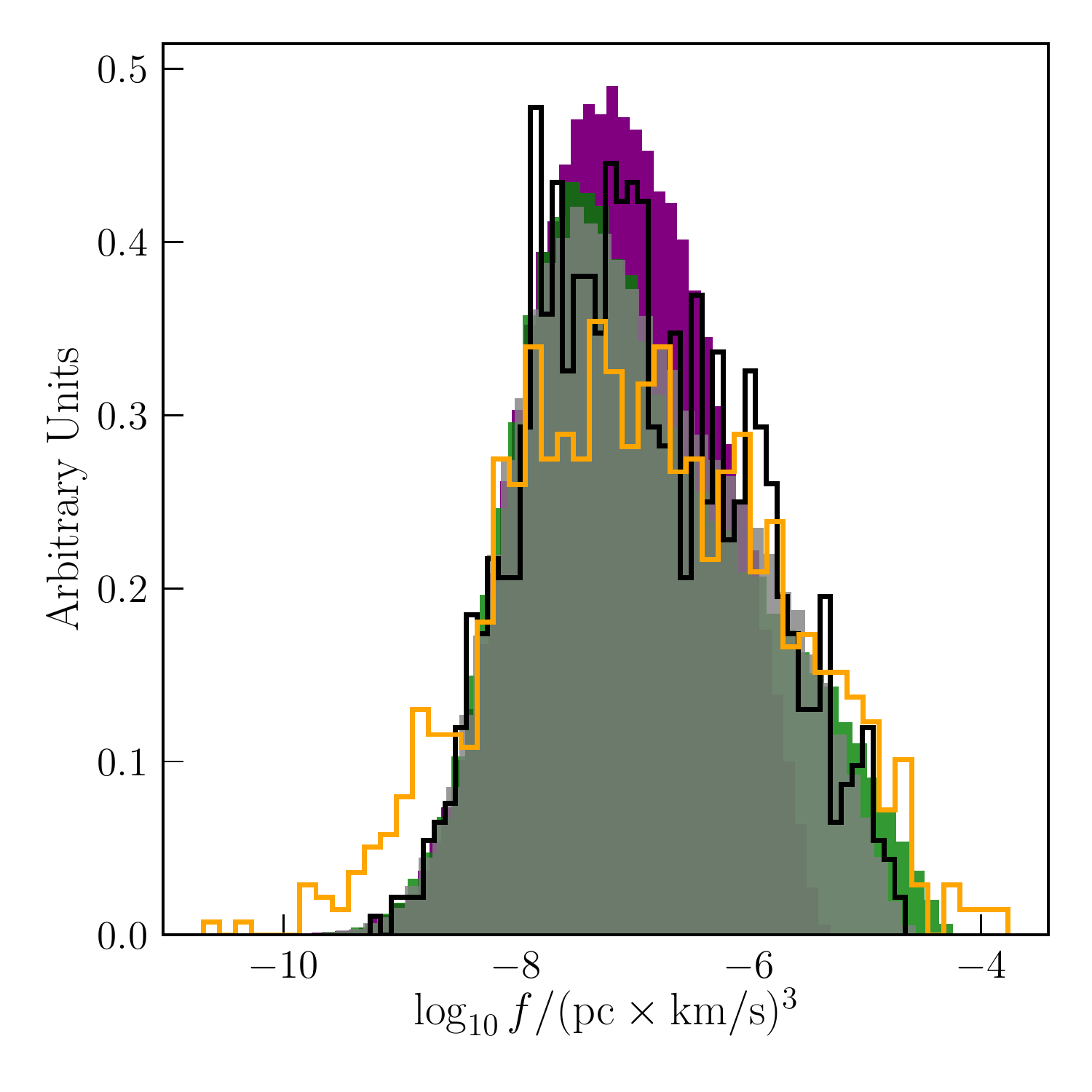}
\caption{Left: Heatmap of the six-dimensional probability density as a function of radial distance $r$ and total speed $v$ for particles in a King profile with $\vec{\xi} = (M,r_0,c)$ parameters given by Eq.~\eqref{eq:M4params}. Center: a histogram of the phase-space densities of 1000 randomly generated stars drawn from a King profile with the parameters of Eq.~\eqref{eq:M4params} $\vec{\xi} = (1.12\times 10^{5}\,M_\odot,0.77~{\rm pc}, 1.65)$ (black line) and the randomized probabilities of these stars (orange line), compared to the probability density function of the King profile (grey solid), a King profile with $\vec{\xi} = (10^{4}\,M_\odot,0.77~{\rm pc}, 1.65)$ (red solid), and one with $\vec{\xi} = (10^{6}\,M_\odot,0.77~{\rm pc}, 1.65)$ (blue solid). Right: The same probability distribution of stars drawn from the King  profile with  $\vec{\xi} = (1.12\times 10^{5}\,M_\odot,0.77~{\rm pc}, 1.65)$ (black and orange lines, grey solid), compared to the distribution of globular clusters with $\vec{\xi} = (1.12\times 10^{5}\,M_\odot,0.5~{\rm pc}, 1.84)$ (green solid), and $\vec{\xi} = (1.12\times 10^{5}\,M_\odot,2~{\rm pc}, 1.24)$ (purple solid). The $c$ parameters are chosen to keep $r_t = 34.9$~pc as $r_0$ is varied, and the underlying profiles are constructed using $10^5$ random samples. \label{fig:6Dprobdensities}}
\end{figure}

We demonstrate this approach using a toy example of a King profile and assuming accurate knowledge of the phase-space density for each tracer star. We generate 1000 stars randomly from a King profile modeled on the parameters of the globular cluster M4 (NGC 6121), as the next section will consider the challenging issue to obtaining density estimates for stars, using {\it Gaia} data of M4 as a test case. M4 is the nearest globular cluster to Earth, at $1.90\pm0.01$~kpc distance~\cite{2015ApJ...808...11N}. Fitting it to a King profile, its density distribution can be described by the parameters \cite{McLaughlin:2006mp}
\begin{equation}
c = 1.65,~r_0 = 0.77~\mbox{pc},r_t=34.9~\mbox{pc},~\sigma = 5.4~\mbox{km/s},~M = (1.12^{+0.17}_{-0.14}) \times 10^5\,M_\odot. \label{eq:M4params}
\end{equation}
In Figure~\ref{fig:6Dprobdensities} we show the probability density for this King profile, as a function of radius and speed of a star particle (this compression from six-dimensional phase space to $r$ and $v$ is possible only due to the spherical symmetry). We also show a histogram of the probability density of 1000 randomly generated stars using the King profile parameters of Eq.~\eqref{eq:M4params}, as compared to the distributions predicted for a selection of cluster masses, King radii $r_0$, and concentrations $c$ (with the latter two parameters chosen so that $r_t$ remains at 34.9~pc). As can be seen, when the parameters of the halo are varied, the distribution of probability densities for stars drawn from the cluster will vary as well, allowing the correct $\vec{\xi}$ parameters to be -- in principle -- estimated by minimizing the log-likelihood, given a sufficient number of tracer stars along with sufficiently accurate measurements of the probability density for each star.

In the next section, we will approach the issue of measuring the probability density of tracer stars in much more detail, including errors, followed by an application to real data. Here we merely demonstrate that the initial $\vec{\xi}$ parameters can be recovered given a reasonable error on the measurement of the probability density. 

\begin{figure}[t]
\includegraphics[width=0.5\columnwidth]{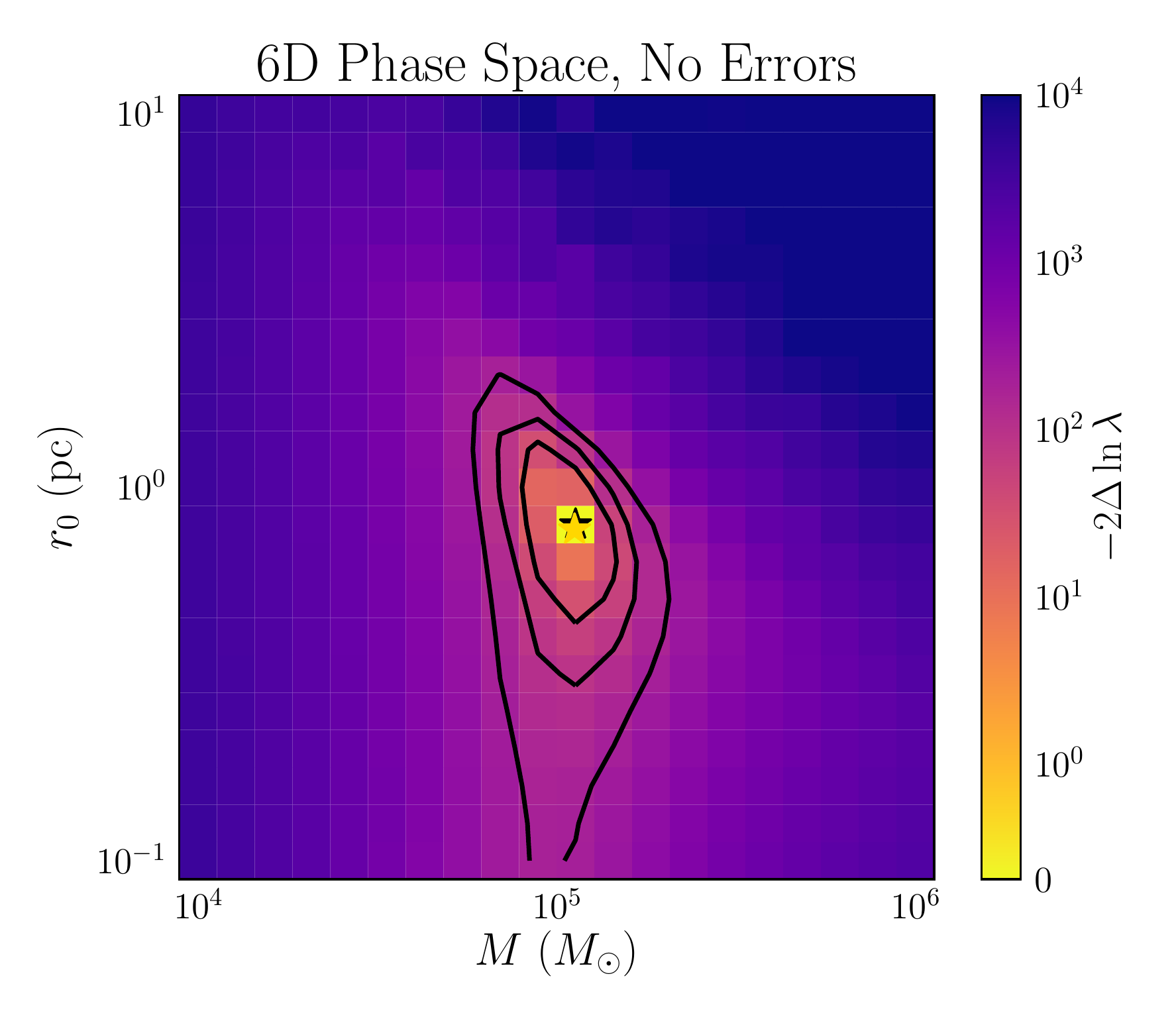}\includegraphics[width=0.5\columnwidth]{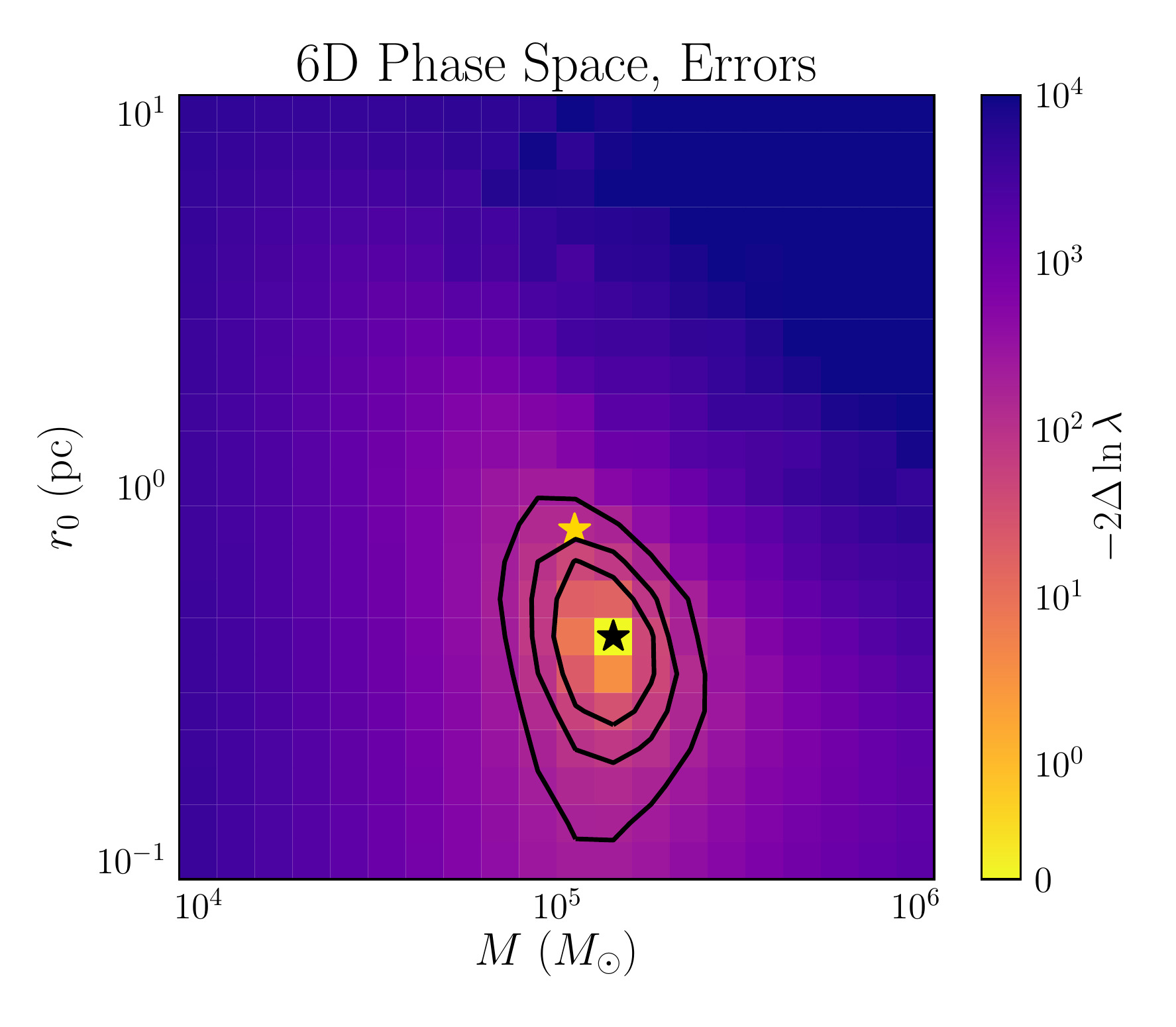}
\caption{Log-likelihood $-2\ln\lambda$ as a function of cluster mass $M$ and King radius $r_0$ (varying the concentration $c$ in order to keep $r_t = 34.9$~pc fixed). The left panel uses 1000 tracer stars assuming perfect knowledge of the probability density values. The right panel randomly perturbs each star's log probability density values by a gaussian of unit width. True value of the mass and King radius is shown with a gold star, the location of the minimum of $-2\ln\lambda$ is indicated with a black star. Contour lines are shown for reference, but do not represent $1\sigma$, $2\sigma$, {\it etc} ellipses.  \label{fig:6dreconstruction}}
\end{figure}

First, we imagine that the measured phase-space density comes from stars with no measurement error. Minimizing $-2\ln \lambda(\vec{\xi})$, varying the mass and the combination of $r_0$  and $c$ while keeping $r_t$ fixed, in the left panel of Figure~\ref{fig:6dreconstruction} we show the log-likelihood as a function of $M$ and $r_0$, for each star setting $[f_{i,\rm low},f_{i,\rm high}] = [0.5f_i,2f_i]$. Extracting the errors using the Fisher matrix, the best fit is
\begin{equation}
M = (1.12\pm 0.03)\times 10^5,M_\odot,~r_0 = 0.79\pm 0.04~{\rm pc}.
\end{equation}
Which exactly matches the input parameters of the system. Note that the small errors here (and throughout the rest of this paper) are statistical only.

Next, to simulate the effects of inaccurate measurements, we wish to perturb each star's phase-space density by some random number. As we will see shortly when realistic errors are considered, the measured density can shift by an order of magnitude or more from the true value. To provide a demonstration of the result of such large shifts on the phase-space density distribution without having a data set with real {\it Gaia} errors, we introduce a random shift to the log of the phase-space densities.

Specifically, for each star, we randomly perturb the true probability density in log space, adding to each log-density a random number drawn from a gaussian of unit variance centered at zero. The resulting distribution of $f$ is shown in orange in Figure~\ref{fig:6Dprobdensities}. We then repeat the fitting procedure described for the non-perturbed densities. The results are shown in the right-hand panel of Figure~\ref{fig:6dreconstruction}, with best-fit mass and radius
\begin{equation}
M = (1.44\pm0.04)\times 10^5\,M_\odot,~r_0 = 0.40\pm0.02~{\rm pc}.
\end{equation}
We again emphasize that this random perturbation of densities is provided only for illustrative purposes, and is not intended to reproduce the complete effects of realistic errors in the data.

Looking at the distributions in Figure~\ref{fig:6Dprobdensities}, it is understandable why introducing the errors in this way reconstructs the mass relatively accurately, but not $r_0$. Adding a random but unbiased ``mismeasurement'' of the probability density widens the distribution, while not shifting the average $f$. As can be seen in Figure~\ref{fig:6Dprobdensities}, decreasing the mass moves the distribution of $f$ to higher values, while decreasing $r_0$ widens the distribution while not significantly changing the mean. Thus, our simple mismeasurement assumption will systematically lower $r_0$ while not having a significant impact on $M$. Accurate measurement (or understanding of the bias introduced by errors) will therefore be critical to reconstructing the density profile. However, the exact effects of these errors (and the best methods to correct for them) requires a better model of the effect of errors on the phase-space distribution than the very crude addition of a random shift to the log-densities, as was done in this example.

\section{Entropy from Measurement Errors \label{sec:mismeasure}}

Having shown that the (conserved) phase-space density can provide insight into the masses of systems that were once gravitationally bound, we now turn to the practical realities of applying this idea to real data. In this Section, we concentrate on the effects of {\it Gaia} measurement errors in position and proper motion on the determination of the phase-space density. These measurement errors serve to artificially increase the spatial extent and velocity dispersion of the measured system: in effect a collection of stars will appear bigger and ``hotter'' than it is in reality. In addition, membership of any cluster cannot be assigned perfectly, and there will always be foreground stars that are moving much faster at a given radius than an equivalent bound star. 

However, the fact that these observational errors tend to move stars to points in phase space with lower density (and higher volumes) provides a method to identify and correct for these errors. We will demonstrate techniques to subtract off this ``entropy injection'' using the M4 globular cluster as observed in {\it Gaia}. We should emphasize that the particular cuts used here are tuned to the cluster in question, but we believe the approach could -- with further work -- be applied with more robust statistical methods to more general systems.

As discussed in Section~\ref{sec:volume}, M4 is the nearest globular cluster to Earth. Unfortunately, despite this proximity, radial distances and motions for most stars at distances of $\sim 2$~kpc are not well measured by {\it Gaia} in DR2. Therefore, we will have to compress our phase space from six dimensions to four (corresponding to angles and angular velocities on the sky). Since the cluster is spherically symmetric, this is not a significant loss of information, but disrupted systems would require additional observational input to determine distance and radial motion. 

In general, the four-dimensional phase space volume is not conserved by Liouville's theorem; for systems that lack the spherical symmetry, other techniques would have to be used to estimate full six-volume from the {\it Gaia} measurements. In this highly symmetric situation however, we can integrate the six-dimensional phase-space density of the King profile Eq.~\eqref{eq:king} along the $z$ and $v_z$ directions to calculate a probability for a set of four-dimensional densities measured from some set of tracer stars.

\medskip

\begin{figure}[th]
\includegraphics[width=0.4\columnwidth]{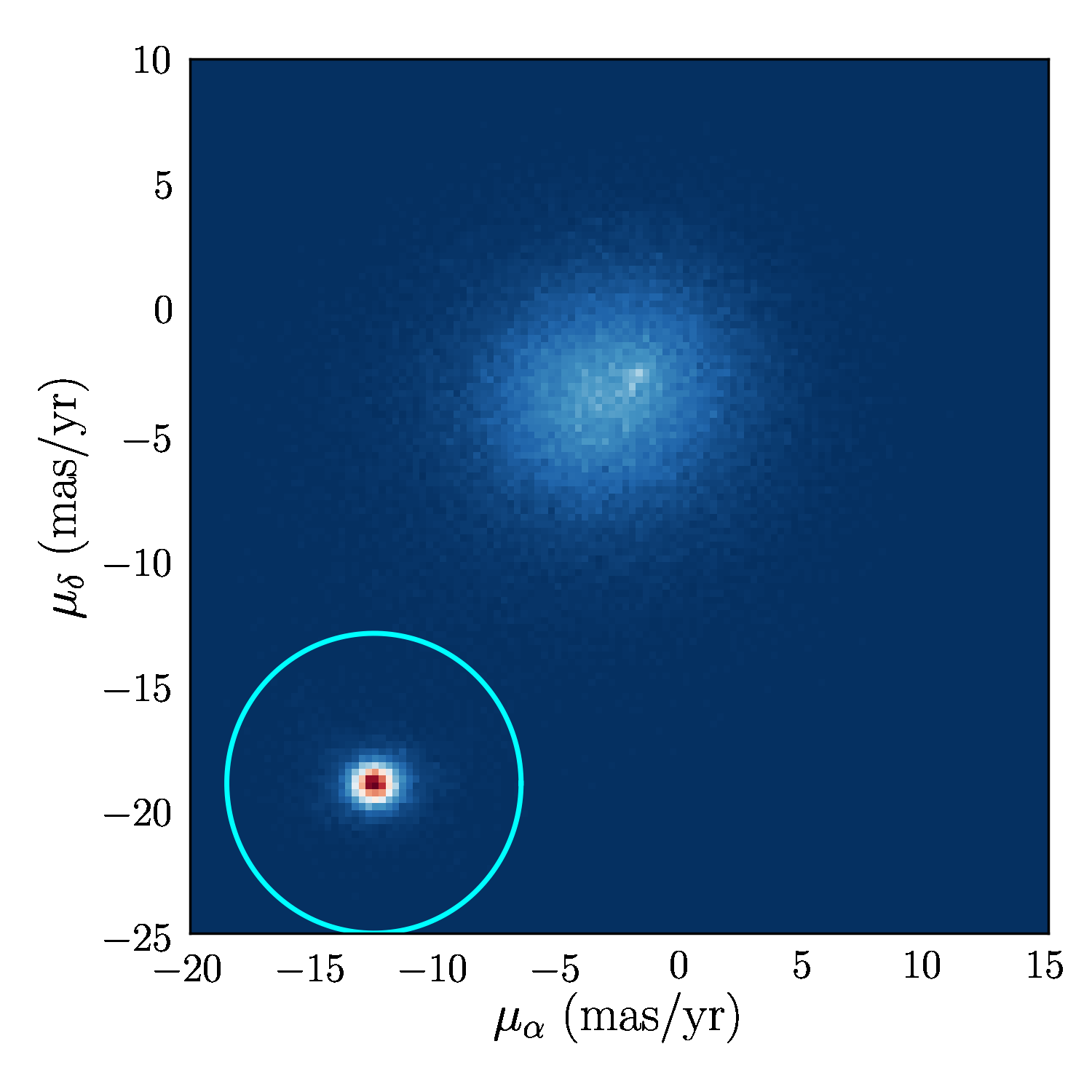}
\caption{Proper motions of all stars in {\em Gaia} DR2 within $1^\circ$ of the center of the M4 globular cluster. The cluster corresponds to the overdensity in the lower left. The proper motion cut is shown by the cyan circle. 
\label{fig:M4pm}}
\end{figure}

However, estimating phase-space density in this sample is nontrivial even in the reduced number of dimensions, due to measurement errors, contamination in the sample, and completeness issues.
 Selecting stars in the {\it Gaia} DR2 catalogue within $1^\circ$ of the center of M4 yields foreground stars which are not part of the cluster, as well as those that are gravitationally bound. Most of the former category (though not all) can be removed by cutting on the proper motions of the stars.\footnote{The data was retrieved from the {\it Gaia} repository \url{http://gea.esac.esa.int} using the ADQL query: \newline \texttt{SELECT source\_id,ra,ra\_error,dec,dec\_error,parallax,parallax\_error,phot\_g\_mean\_mag,bp\_rp,radial\_velocity, radial\_velocity\_error,
 phot\_variable\_flag,teff\_val,a\_g\_val,pmra,pmdec,pmra\_error,pmdec\_error FROM gaiadr2.gaia\_source  
 WHERE CONTAINS(POINT('ICRS',gaiadr2.gaia\_source.ra,gaiadr2.gaia\_source.dec),
\newline CIRCLE('ICRS',COORD1(EPOCH\_PROP\_POS(245.89675000000003,-26.52575,0,-17.6300,-21.5700,70.4000,2000,2015.5)),
 COORD2(EPOCH\_PROP\_POS(245.89675000000003,-26.52575,0,-17.6300,-21.5700,70.4000,2000,2015.5)),1.0))=1    AND  (parallax<=0.75)}} In Figure~\ref{fig:M4pm}, the overdensity corresponding to M4 can clearly be seen in proper motion space. We select stars with a proper motion $\mu < 6$~mas/yr of the center of this overdensity (corresponding to $\mu_\alpha = -12.5$, $\mu_\delta = -18.98$~mas/yr), and with parallaxes less than $0.75$~mas (corresponding to distances greater than $1.3$~kpc). We do not attempt a parallax cut to narrow the stellar distance down to a region around M4, given the large parallax errors for stars in M4. After these cuts, 20,919 stars remain in the sample.

To see the effect of errors on the phase-space density, we create a simulated M4 cluster with realistic {\it Gaia} measurement errors. We generate 20,000 simulated stars from a King profile using the best-fit parameters for M4, Eq.~\eqref{eq:M4params}. Foreground stars are simulated by taking stars in an annulus from $1^\circ$ to $2^\circ$ around the center of M4 in the data, and applying the same proper motion and parallax cuts as for the cluster itself. As the annulus has an angular area three times as large as the M4 signal region, one third of the annulus stars are randomly selected as foreground stars and moved to a random location within the simulated cluster. This results in 542 additional stars, to be added to the 20,000 simulated stars from the King profile. 

\begin{figure}[t]
\includegraphics[width=\columnwidth]{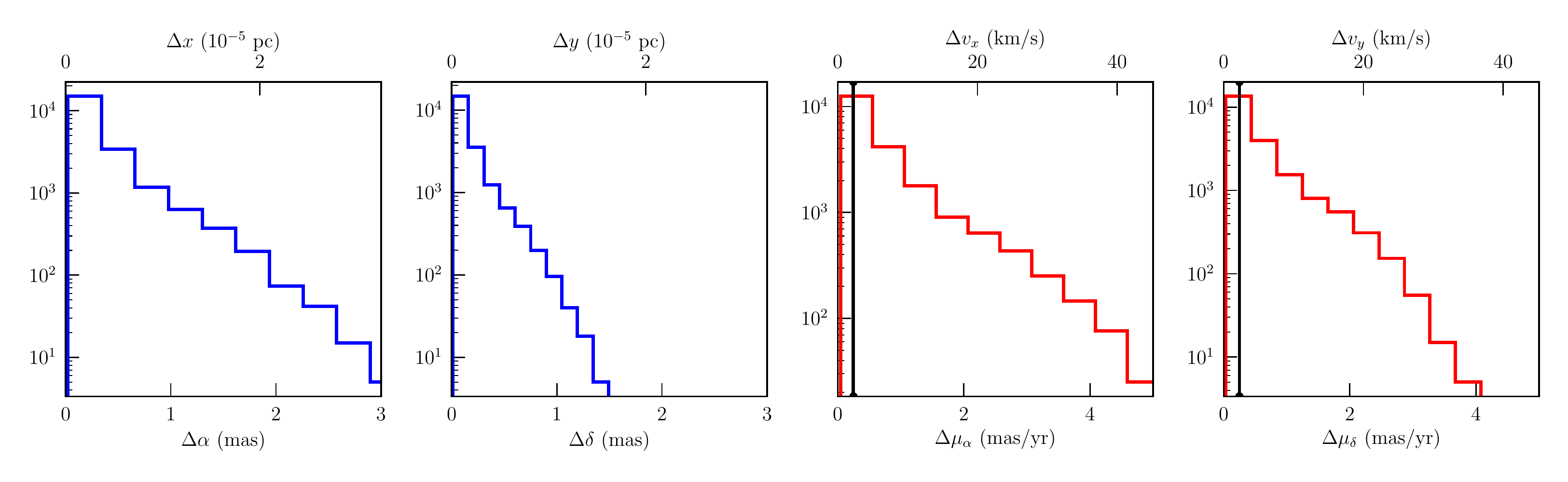}
\caption{Measurement errors reported by {\it Gaia} for stars in the $1^\circ$ centered on the location of M4 and within 6.0 mas/yr of the average proper motion of the cluster. Errors are reported in both angular location (ra $\alpha$ and dec $\delta$) and proper motion ($\mu_\alpha$ and $\mu_\delta$), as well as physical location ($x$ and $y$) and velocity components ($v_x$ and $v_y$), assuming a distance to M4 of $1.904$~kpc. \label{fig:gaia_errors}}
\end{figure}

We simulate realistic {\it Gaia} measurement errors by randomly assigning each simulated cluster and foreground star the estimated errors in position and proper motion of a real star in the {\it Gaia} M4 sample, see Figure~\ref{fig:gaia_errors}. We then perturb each star's location in phase space, by drawing random numbers from a Gaussian distribution centered on each of $x_i$ and $v_i$, using the star's assigned error in that dimension for the standard deviation. We note that one obvious weakness of this method to produce ``realistic'' errors is that it removes any spatial correlations that might be present in the actual errors. In general, these are expected to be small in the {\it Gaia} data (and decrease as the mission progresses) \cite{2012A&A...543A..15H}, though as we will see, the density of stars in M4 does introduce additional density mismeasurement correlated with distance from the cluster's center.

We measure the phase-space density using the program \textsc{EnBiD} \cite{2006MNRAS.373.1293S}, which numerically estimates the phase-space density of a discrete sample in arbitrary dimensions by recursively subdividing the sample into roughly equal sets. 
For systems with nearly-spherical symmetry, such as the M4 cluster, we use a spherical kernel, while for streams (considered in the next Section), we use an adaptive metric. In each case we consider, we tune the \textsc{EnBiD} kernel parameters so that the resulting density estimates match as closely as possible the analytic result in simulated data.

\begin{figure}[t]
\includegraphics[width=0.19\columnwidth]{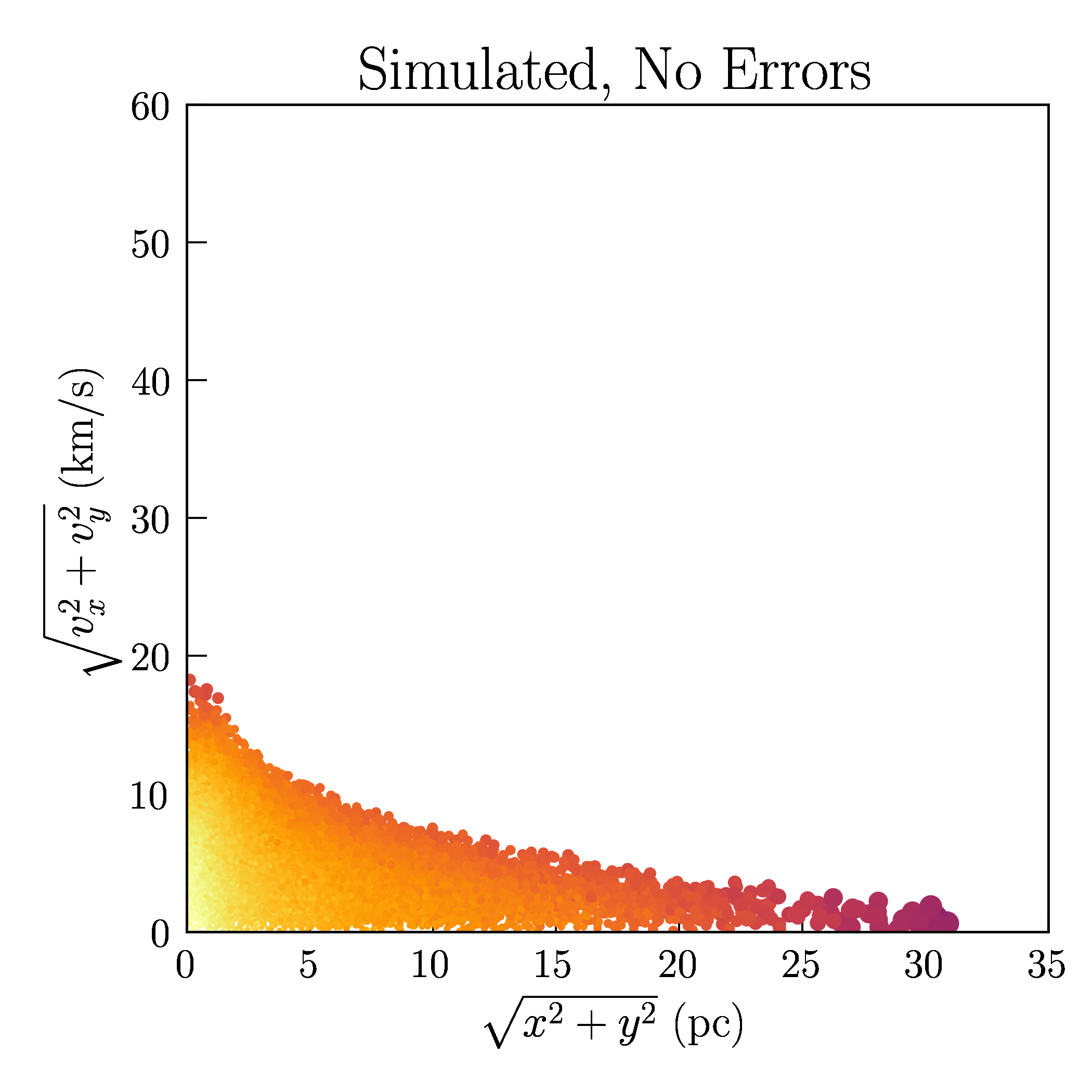}\includegraphics[width=0.19\columnwidth]{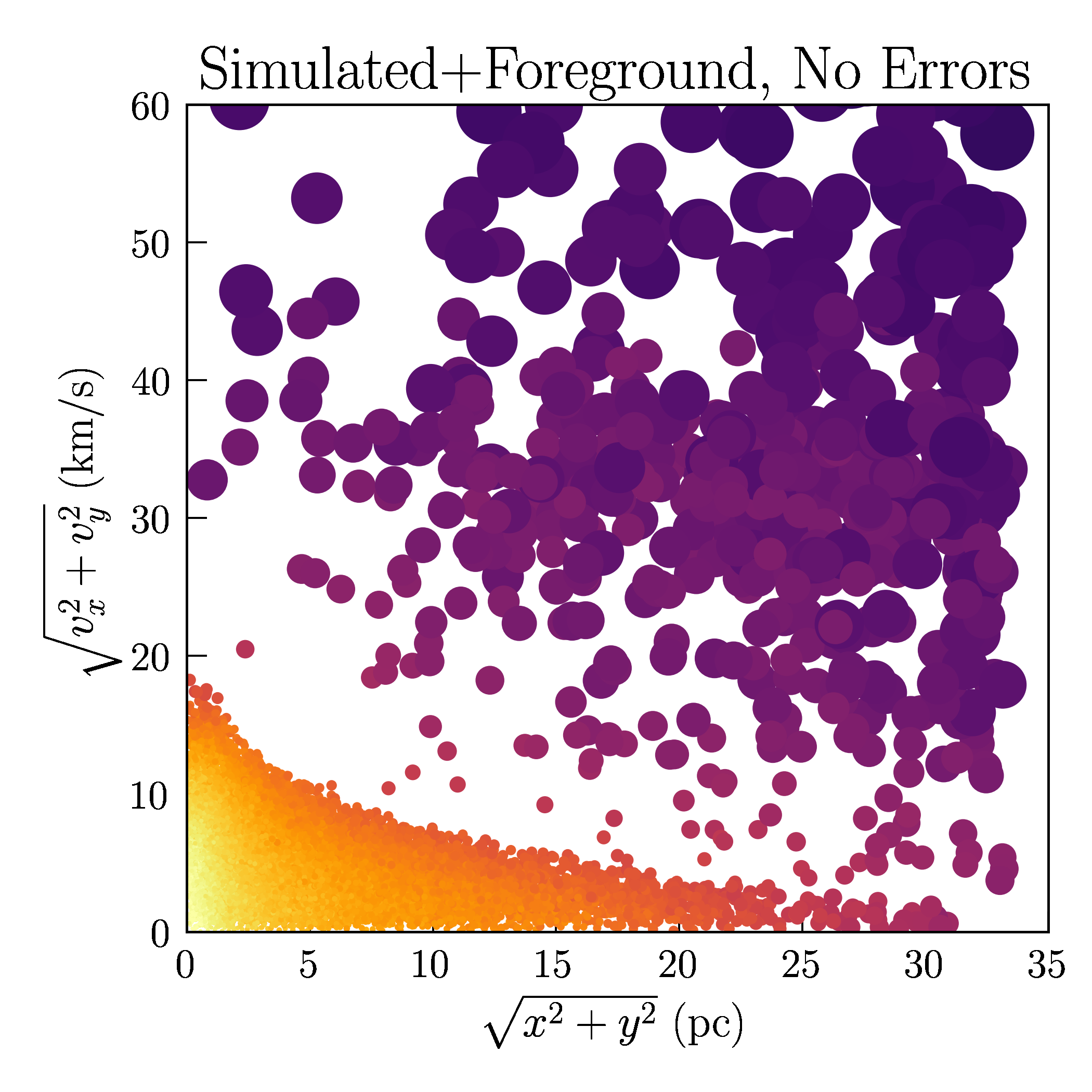}\includegraphics[width=0.19\columnwidth]{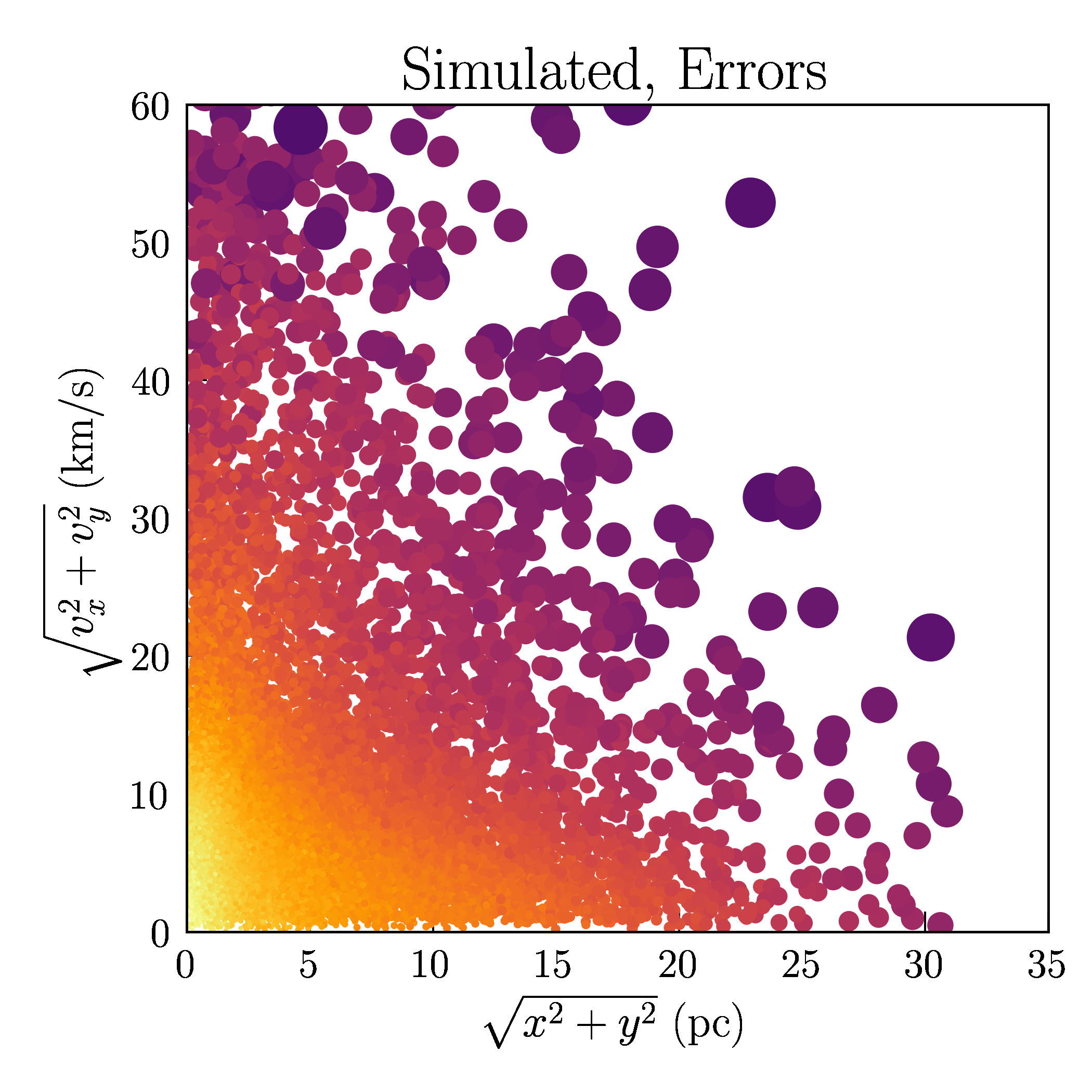}\includegraphics[width=0.19\columnwidth]{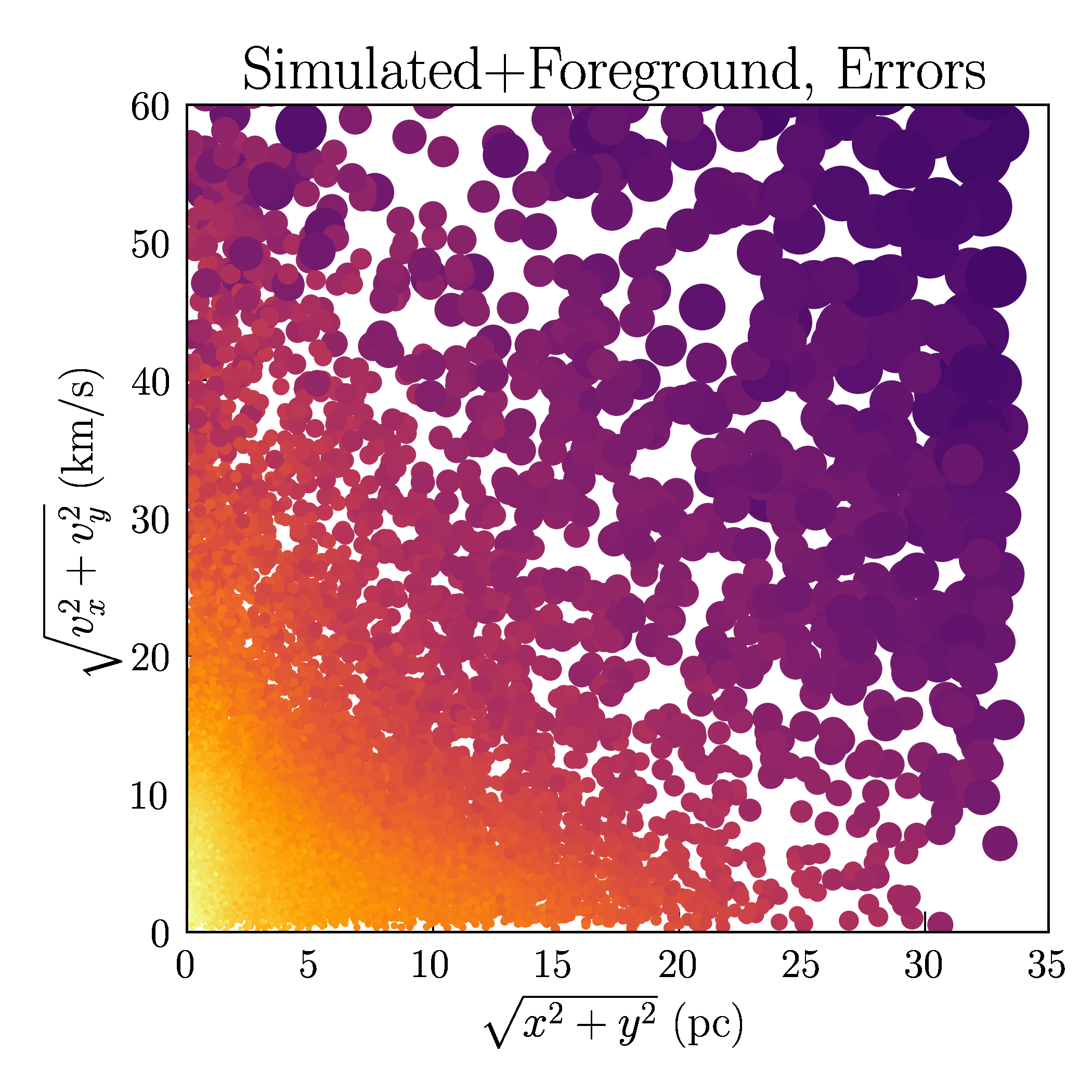}\includegraphics[width=0.225\columnwidth]{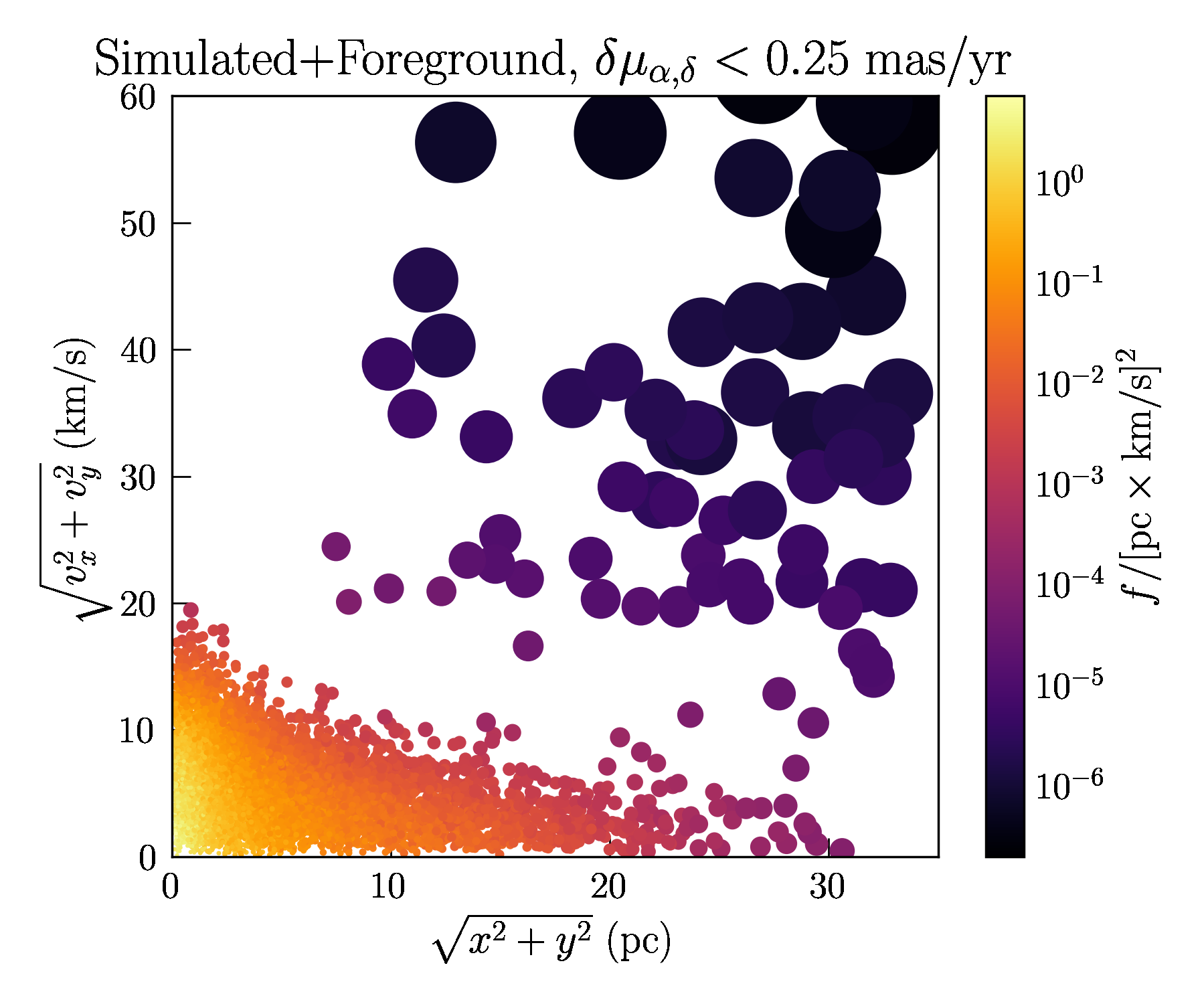}
\caption{Scatter plot of the radial position and speed relative to the average motion of the 20,000 simulated M4 stars, without errors and foreground stars (far left), without errors and with 542 foreground stars (center left), with errors (center), with errors and foreground (center right), and after the cut on proper motion errors Eq.~\eqref{eq:pmcut} (far right). The size and color of each star indicate the relative phase-space density of each star (darker and larger circles indicating lower density/greater phase-space volume).
\label{fig:simulated_scatterplot}}
\end{figure}

\begin{figure}[t]
\includegraphics[width=0.4\columnwidth]{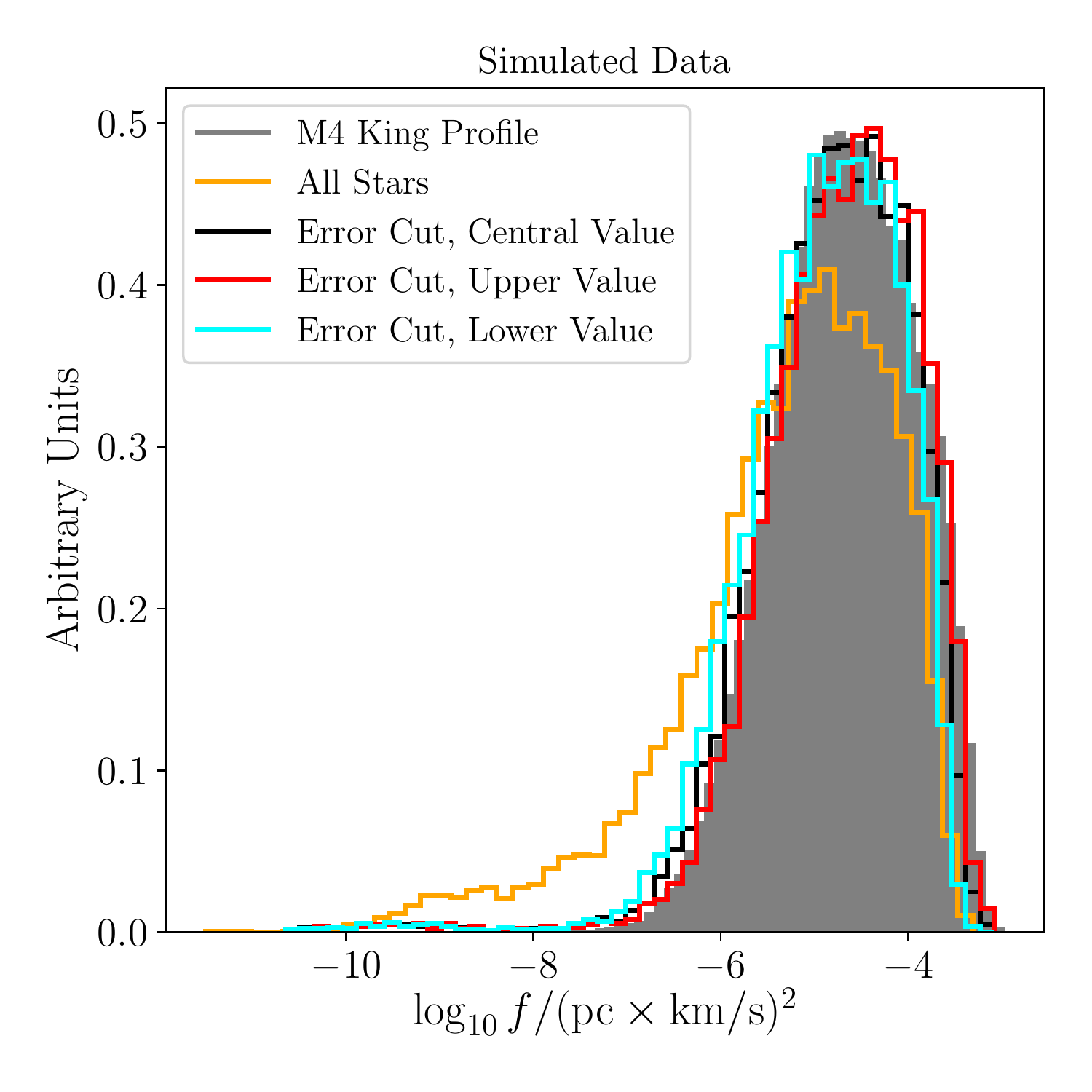}\includegraphics[width=0.4\columnwidth]{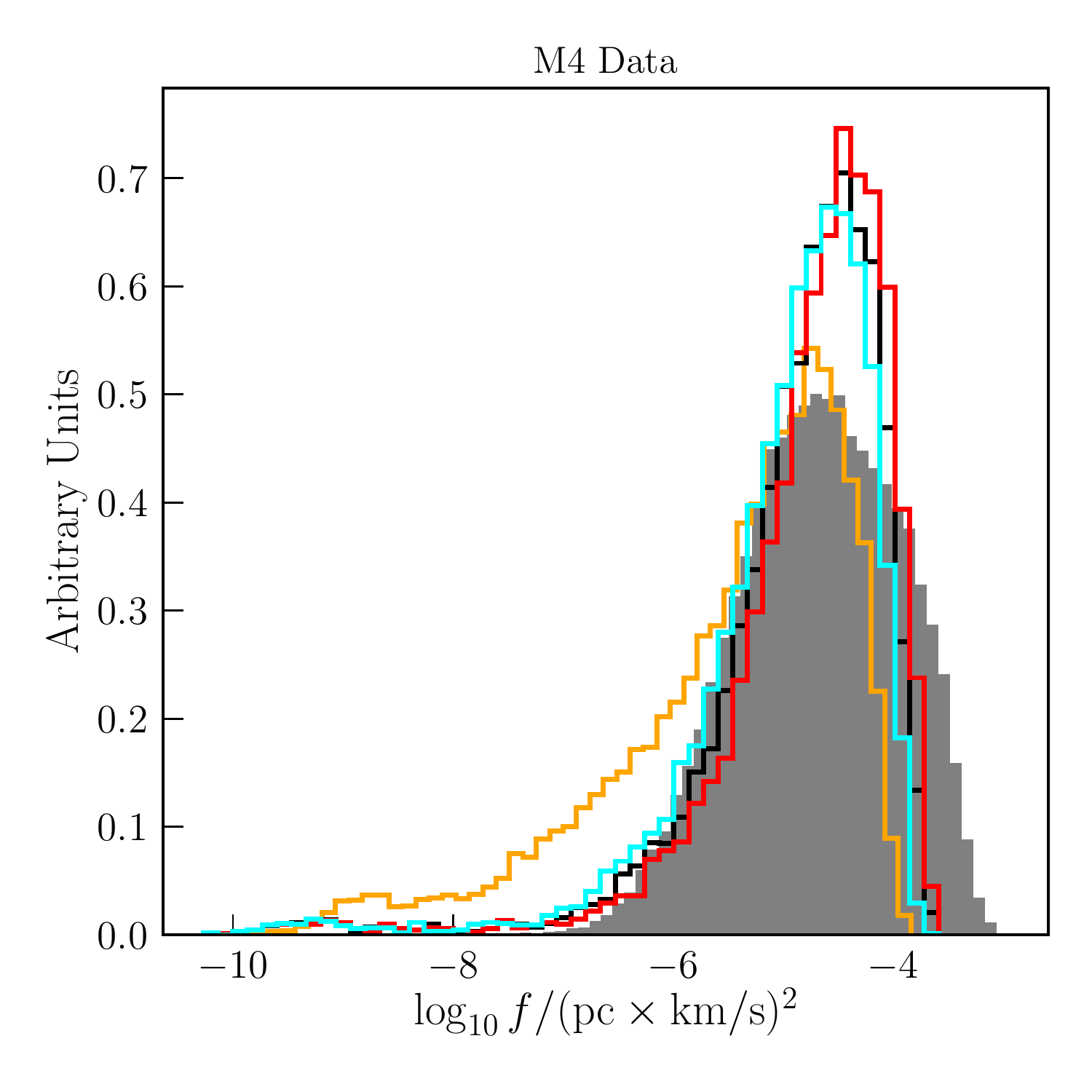}
\caption{Histograms of the phase-space density of the stars. Left: simulated stars drawn from a King profile with the parameters of Eq.~\eqref{eq:M4params} (shown in grey here). Right: distribution of stars in {\it Gaia} data. The distribution of all stars, including measurement errors, is shown in orange. The phase-space density distribution of stars with small proper motion errors Eq.~\eqref{eq:pmcut} is shown in black. The upper and lower range of the phase-space density for each low-error star, extracted from the proper motion error ellipse, are shown in red and cyan, respectively.
\label{fig:4Dprobdensities}}
\end{figure}

In Figure~\ref{fig:simulated_scatterplot}, we show the distribution of simulated stars in $r$ and $v$, color- and size-coded by phase-space density. There are two effects of interest here. First, without applying errors, the foreground stars are easy to distinguish from the gravitationally-bound stars: at a given radius, the foreground stars are moving much faster relative to the frame of reference of M4's center of mass. Applying realistic errors erases this contrast. Second, the effect of measurement errors on the positions and velocities moves many stars in the cluster out to the lower tail of the density distribution. This can be seen in the 3$^{\rm rd}$ and $4^{\rm th}$ panels of Figure~\ref{fig:simulated_scatterplot}, where the number of stars with low phase-space densities has clearly increased over the true distribution. This can also be seen in Figure~\ref{fig:4Dprobdensities} by comparing the phase-space density distribution of the full set of stars in four dimensions to that of the King profile with the parameters of M4. 

\begin{figure}[t]
\includegraphics[width=0.33\columnwidth]{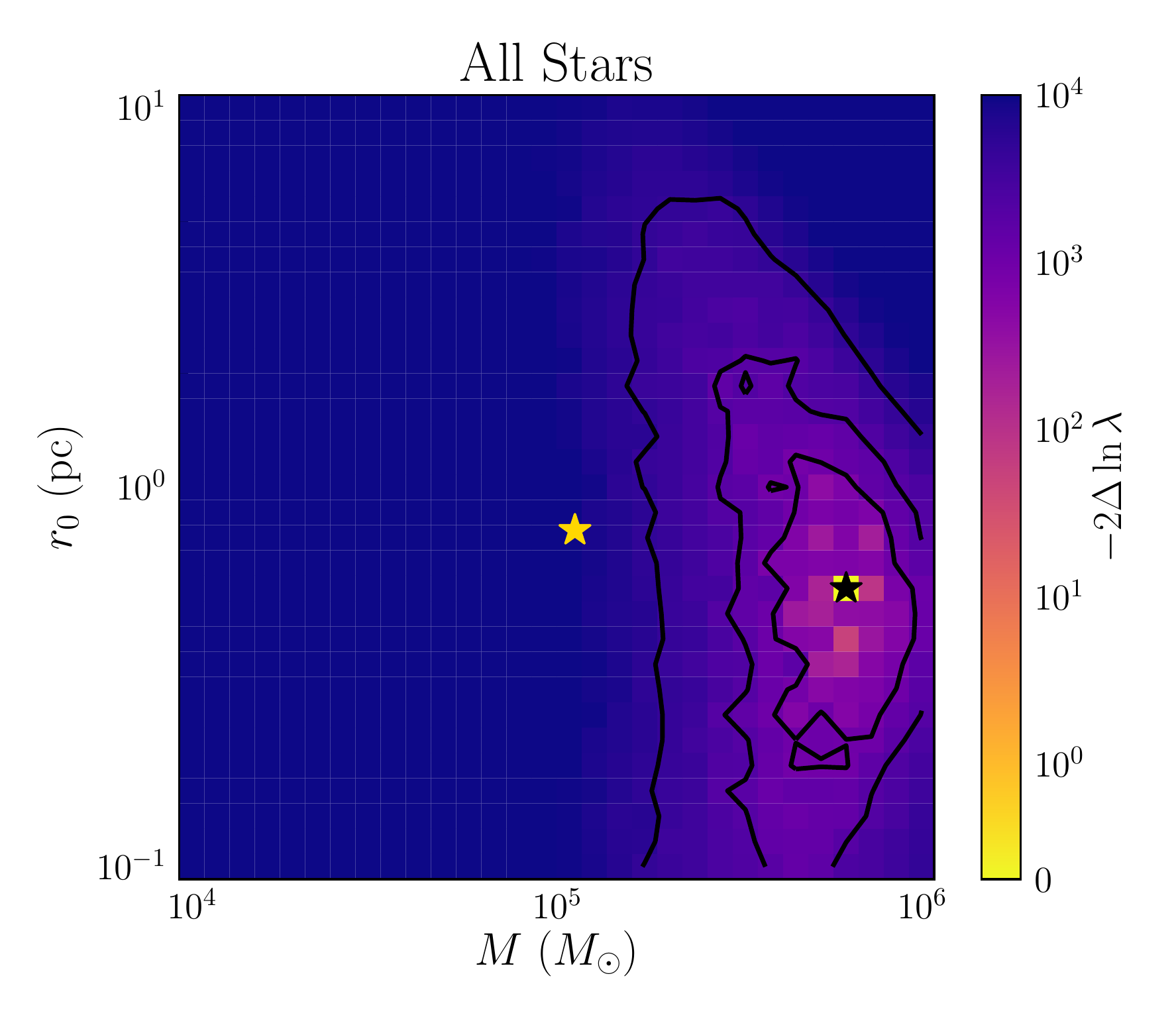}\includegraphics[width=0.33\columnwidth]{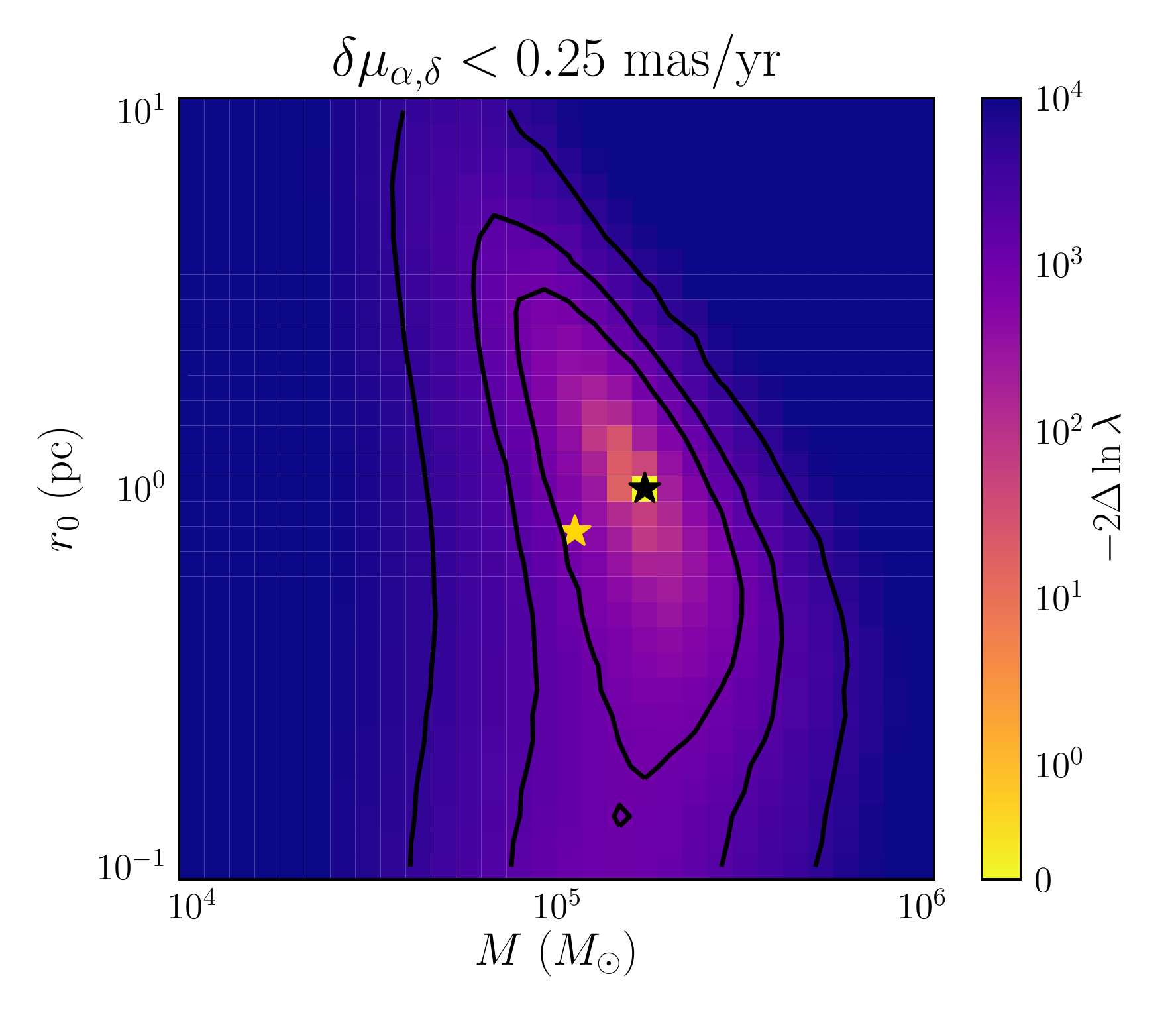}\includegraphics[width=0.33\columnwidth]{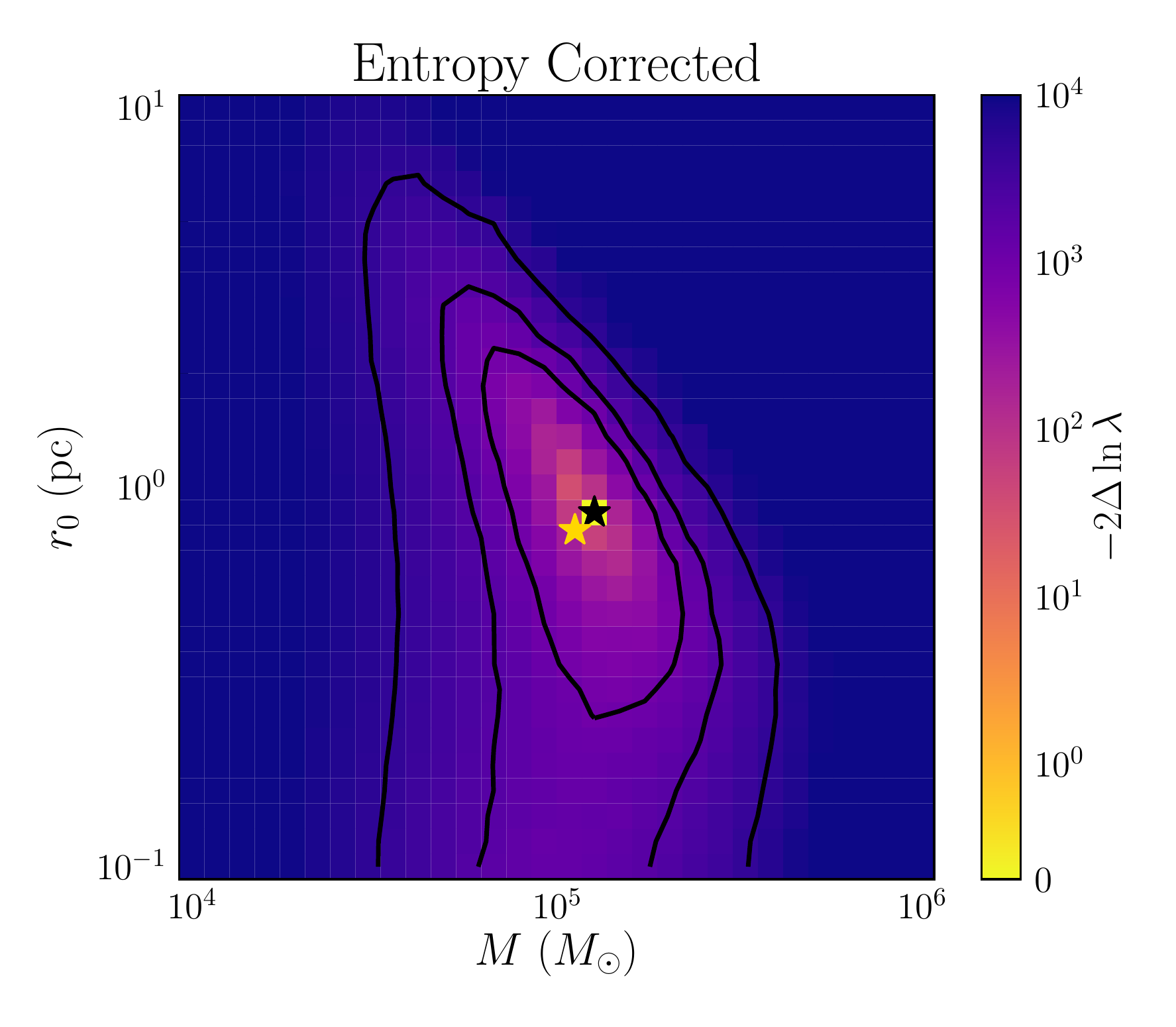}
\caption{Log-likelihood $-2\ln\lambda$ as a function of cluster mass $M$ and King radius $r_0$ (varying the concentration $c$ in order to keep $r_t = 34.9$~pc fixed) for a simulated globular cluster using four-dimensional phase space. The left panel uses all 20,000 simulated cluster stars and 542 foreground stars, with gaussian errors applied to their position and velocity. The center panel places a cut on the proper motion errors of $\delta \mu_{\alpha,\delta} < 0.25$~mas/yr. The right panel additionally corrects for the entropy injection of measurement errors by taking the phase-space density range for each star to be the central value and the average of upper excursions within the $1\sigma$ variation in the proper motion error ellipse, along with a cut of $f>10^{-7}~[{\rm pc \times km/s}]^{-2}$.
The true value of the mass and King radius is shown with a gold star, the location of the minimum of $-2\ln\lambda$ is indicated with a black star. Contour lines are shown for reference, but do not represent $1\sigma$, $2\sigma$, {\it etc} ellipses. 
\label{fig:simulated_mass_r_fits}}
\end{figure}

The observation that errors cause stars to ``move'' to preferentially lower density, rather than equally to higher or lower, can be understood as the measurement errors acting to ``heat'' the cluster, injecting entropy and therefore increasing the phase-space volume (and decreasing the average phase-space density). In the left panel of Figure~\ref{fig:simulated_mass_r_fits}, we show the mass and King radius one would extract for the simulated M4 cluster using all 20,000 cluster stars and 542 foreground stars. Using the range $[f_{i, \rm low},f_{i, \rm high}] = [0.1f_i,10f_i]$ for each star, the resulting best fit values are
\begin{equation}
\mbox{Simulated, all stars}:\quad M= (6.16 \pm 0.06)\times 10^5\,M_\odot,~r_0 = 0.50\pm0.01~{\rm pc}.
\end{equation}
The mass in particular is off by nearly an order of magnitude. Some method is required to reduce the entropy injection of the measurement errors. Keep in mind that the specific cuts we develop here are designed with the M4 system in mind; however they can be adapted and generalized to other systems.

As a first pass, we attempt to mitigate the effects of the measurement errors by cutting those stars where the reported error is exceptionally large (note that {\it Gaia} reported errors are largely a function of the effective number of CCD crossings for the star \cite{LL:GAIA-JDB-022}). As can be seen in Figure~\ref{fig:gaia_errors}, the positional errors are small compared to the radius of M4 (35~pc), and so are not the main source of entropy. The proper motion errors, however, can easily add tens of km/s of velocity, in a system where the escape velocity is no more than 15~km/s. We therefore apply a first cut, requiring
\begin{equation}
|\delta\mu_{\alpha}|~\mbox{or }|\delta\mu_{\delta}| <0.25~\mbox{mas/yr}. \label{eq:pmcut}
\end{equation}
The measurement error cut drops the number of simulated stars to 8,725: 8,649 stars from the cluster and 76 foreground stars.\footnote{This procedure will notably affect the completeness of the sample, and can bias slightly the phase-space density measurements if the errors and phase-space locations are correlated, but we leave studying the impact of completeness on the inferred phase-space densities to future work.} The right panel in Figure~\ref{fig:simulated_scatterplot} shows the distribution of these stars in radius and proper speed, as well as their distribution in phase-space density. It should be clear that this cut goes a long way towards reducing the effect of the measurement errors on the phase-space density distribution (see also Figure~\ref{fig:4Dprobdensities}). Further, as can be seen, the foreground stars form an easily-identified tail of the phase-space distribution: a cut requiring $f >10^{-7}~[{\rm pc\times km/s}]^{-2}$ removes nearly every foreground star.

Taking this well-measured set of stars, we can then optimize the log-likelihood of the King model. For each star we calculate a data-driven $[f_{i,\rm low},f_{i,\rm high}]$ by moving each star within its $1\sigma$ proper motion error ellipse and recalculating the phase-space density of the entire set of stars after perturbing this one star.\footnote{Given the relative accuracy between the proper motion and position measurements of {\it Gaia}, we ignore the latter's effect on the phase-space density measurements. In cases where multiple sources of error are competitive, the effect on the phase-space density should be considered by moving the star in all relevant dimensions.} This is then repeated for each star. The average excursion above and below the central measured phase-space density provide an estimate of likely high- and low-values for each star's density, given measurement errors. The scan over $M$ and $r_0$ for those stars with small proper-motion errors is shown in the center panel of Figure~\ref{fig:simulated_mass_r_fits}; the minimum occurring at
\begin{equation}
\mbox{Simulated, $\delta \mu$~cut}:\quad M= (1.83 \pm 0.02)\times 10^5\,M_\odot,~r_0 = 1.00\pm0.02~{\rm pc}.
\end{equation}
Recall that the quoted errors are purely statistical, and is small due to the large number of stars in the sample.

The slightly too-large mass and radius are both a result of the entropy injected by the measurement errors. By taking the central value of the density within the error ellipse of the proper motion, this biases the density distribution towards lower values. Measurement errors tend to {\it decrease} average densities, rather than increase them, therefore the central density for each star is, on average, lower than the actual density before measurement errors. 

To counteract this, rather than taking symmetric errors around the central value of the density, we instead calculate the probability of measuring a phase-space density between the central value and the average upper excursion within the $1\sigma$ error ellipse. We additionally cut the foreground stars by requiring $f >10^{-7}~[{\rm pc\times km/s}]^{-2}$. Again minimizing the log-likelihood (right panel, Figure~\ref{fig:simulated_mass_r_fits}), we find
\begin{equation}
\mbox{Simulated, entropy correction}:\quad M = (1.13 \pm 0.01)\times 10^5\,M_\odot,~r_0 = 1.00\pm0.02~{\rm pc},
\end{equation}
which is very close in mass to the true answer, though still overestimating the King radius.

\begin{figure}[t]
\includegraphics[width=0.35\columnwidth]{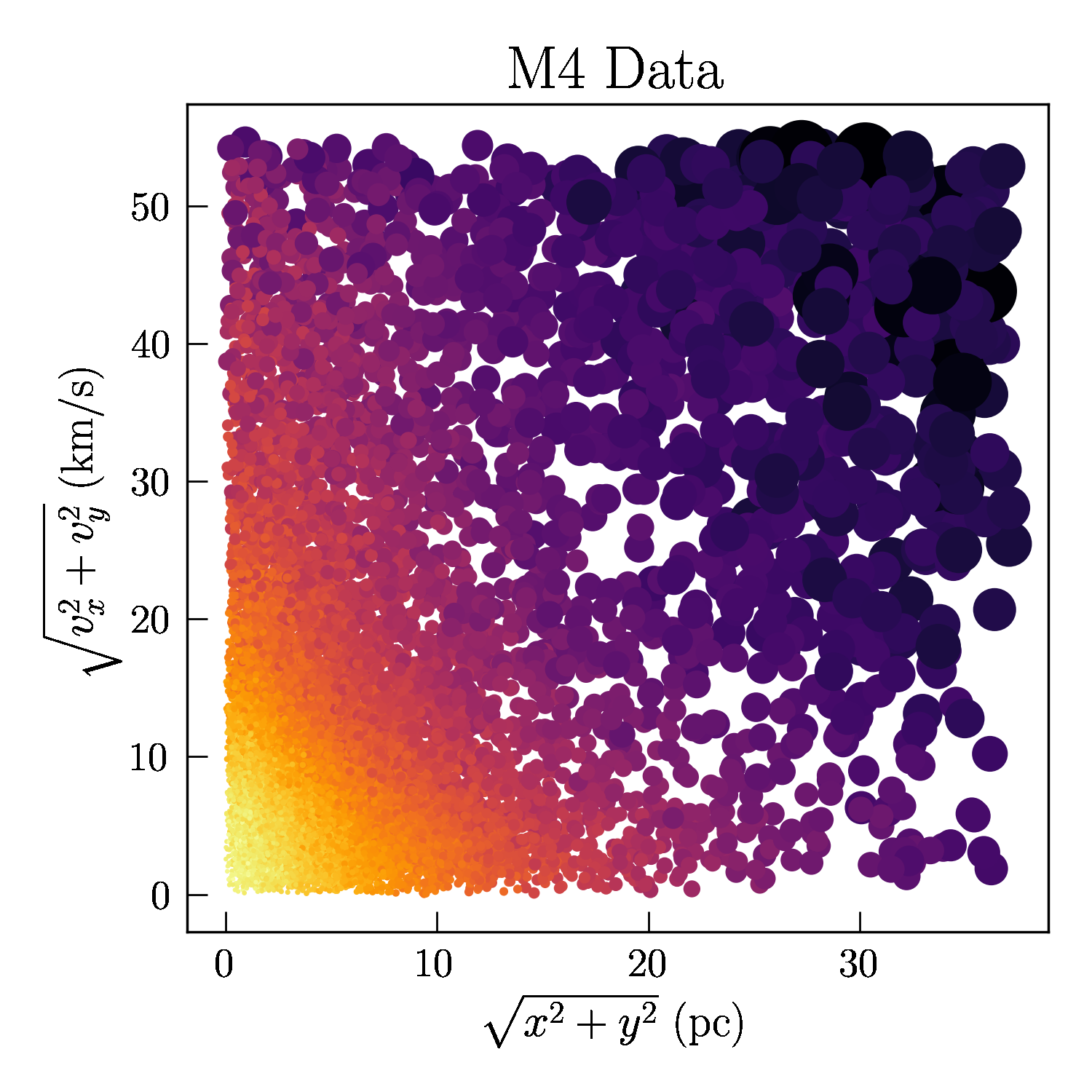}\includegraphics[width=0.42\columnwidth]{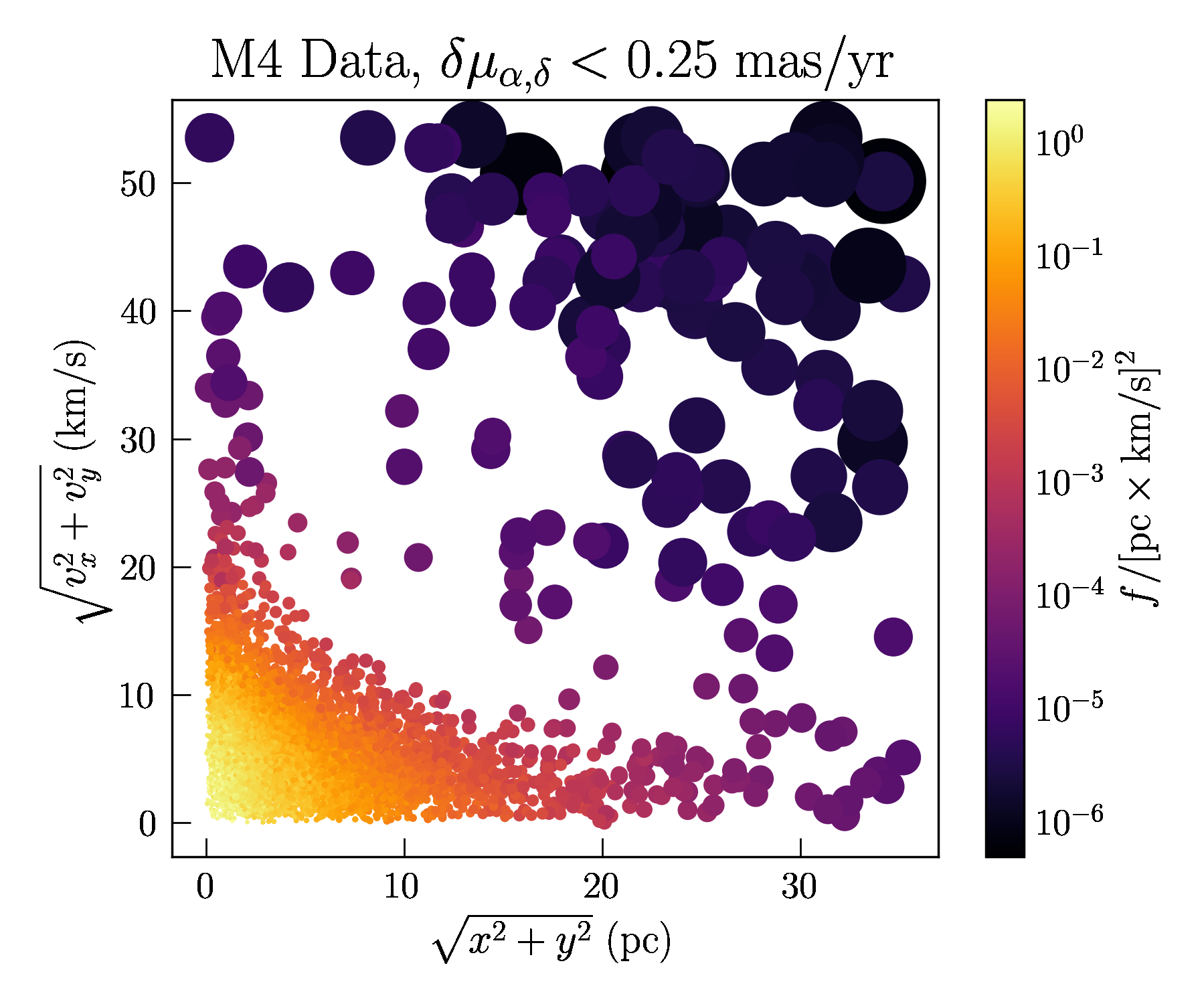}
\caption{Scatter plot of the radial position and speed relative to the average motion of the M4 stars in the {\it Gaia} data without additional selection criteria (left) and after the proper motion cut Eq.~\eqref{eq:pmcut} (right). The size and color of each star indicate the relative contribution of each star to the total phase-space number density.
\label{fig:real_scatterplot}}
\end{figure}

Having developed an analysis strategy based on the simulated cluster, we now apply our methodology to real data. Starting with the 20,919 stars within 6~mas/yr of the center of the M4 proper motion, we apply the error selection criteria of Eq.~\eqref{eq:pmcut}. This drops the number of stars to 9,122 (again, in this work, we ignore the effects of completeness on the inferred phase-space density). The radial position and total proper motion before and after the cuts are shown in Figure~\ref{fig:real_scatterplot}. This should be compared to the equivalent plots in Figure~\ref{fig:simulated_scatterplot}. In particular, even after cuts on proper motion errors, there are notably more stars in the real data with large speeds at low radial distances than there are in the simulated data. This is likely due to crowding of stars at the dense center reducing the measurement precision of {\it Gaia} in a way that is not captured in the estimate of the measurement errors (which only account for the observation history of {\it Gaia} for a particular star and are not correlated with position).

The resulting distributions of the phase-space densities are shown in the right panel of Figure~\ref{fig:4Dprobdensities}, along with the predicted distribution of a King profile with the best-fit mass and King radius from Ref.~\cite{McLaughlin:2006mp}. The crowding effect here serves to truncate the density distribution at high values in a way not captured in our simulation of the cluster. For the densest region of the cluster, the measurement errors -- which do not take into account the local stellar density -- do not accurately reflect the true uncertainty. In addition, the errors in position and velocity measurements in this region are not uncorrelated, reducing the central density far more than a gaussian error assumption would.

\begin{figure}[t]
\includegraphics[width=0.4\columnwidth]{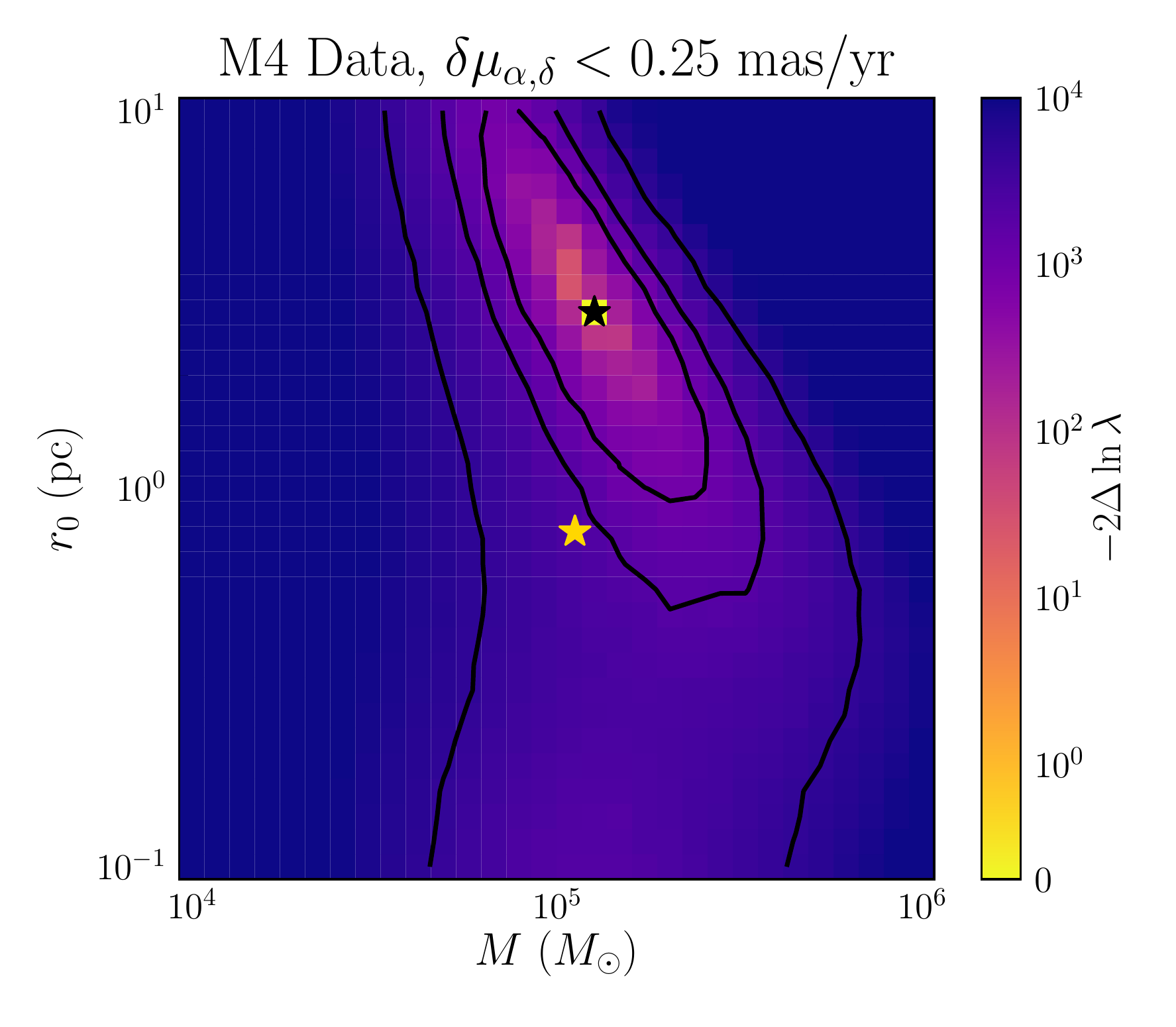}\includegraphics[width=0.4\columnwidth]{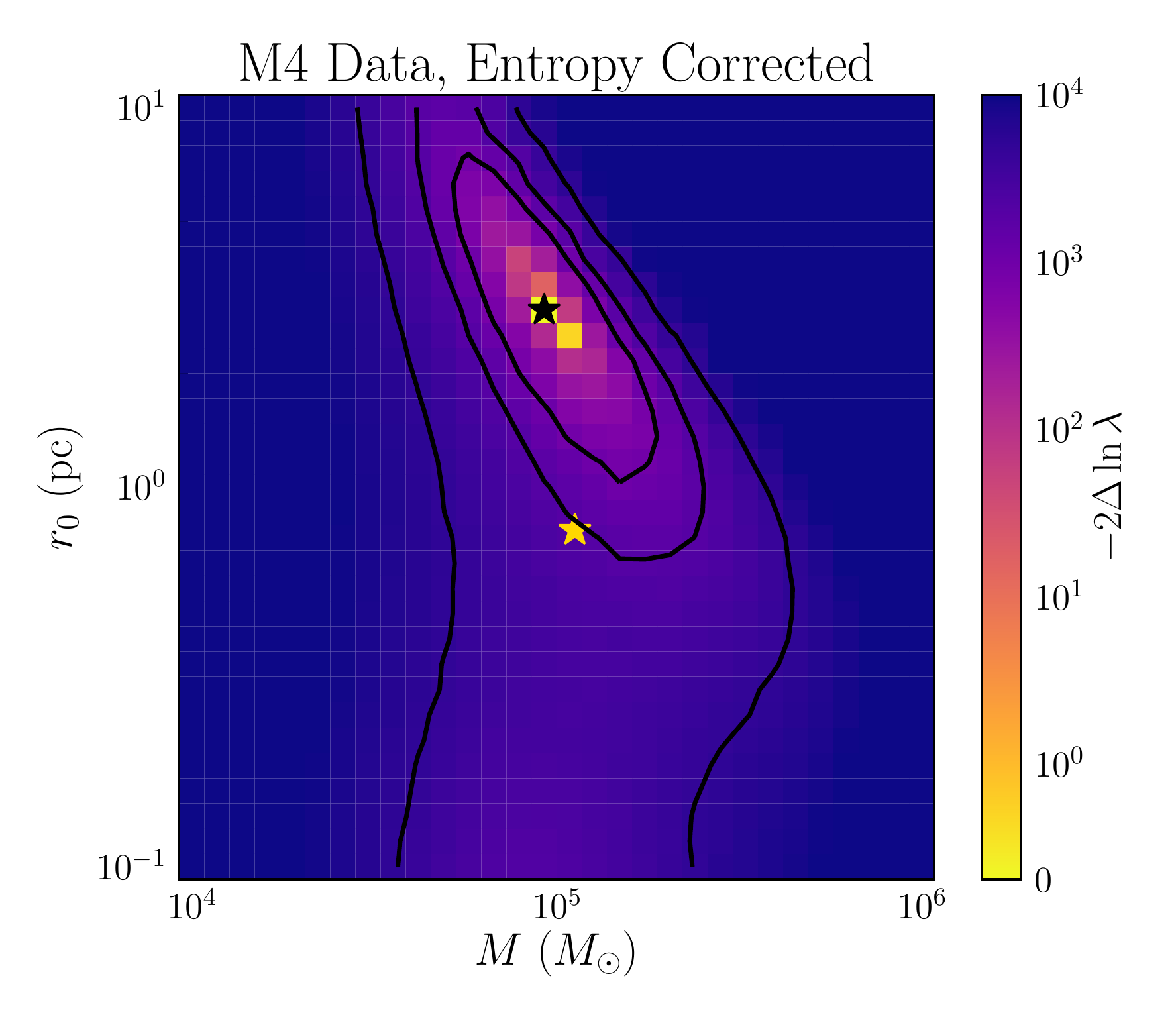}
\caption{Log-likelihood $-2\ln\lambda$ as a function of cluster mass $M$ and King radius $r_0$ (varying the concentration $c$ in order to keep $r_t = 34.9$~pc fixed) for the stars around M4 in the {\it Gaia} catalog using four-dimensional phase space (see text for selection criteria). The left panel requires a cut on the proper motion errors of $\delta \mu_{\alpha,\delta} < 0.25$~mas/yr. The right panel additionally corrects for the entropy injection of measurement errors by taking the phase-space density range for each star to be the central value and the average of upper excursions within the $1\sigma$ variation in the proper motion error ellipse, along with a cut of $f>10^{-7}~[{\rm pc \times km/s}]^{-2}$.
The best-fit values of the mass and King radius from Ref.~\cite{McLaughlin:2006mp} are shown with a gold star, the location of the minimum of $-2\ln\lambda$ is indicated with a black star. Contour lines are shown for reference, but do not represent $1\sigma$, $2\sigma$, {\it etc} ellipses. 
\label{fig:real_mass_r_fits}}
\end{figure}

We then perform the minimization over the log-likelihood, Figure~\ref{fig:real_mass_r_fits}. If we only apply the cut on $\delta\mu_{\alpha,\delta} <0.25$~mas/yr, then the extracted mass and radius are
\begin{equation}
\mbox{{\it Gaia} Data, $\delta\mu$ cut}:\quad M = (1.13 \pm 0.01)\times 10^5\,M_\odot,~r_0 = 3.98\pm0.05~{\rm pc},
\end{equation}
Additionally requiring that $f>10^{-7}~[{\rm pc\times km/s}]^{-2}$ and correcting for the entropy introduced by measurement errors by calculating the probability between the central phase-space density and the average upper excursion within the $1\sigma$ proper motion error ellipse, we calculate our best estimate of the mass and radius
\begin{equation}
\mbox{{\it Gaia} Data, entropy correction}:\quad M = (0.89 \pm 0.01)\times 10^5\,M_\odot,~r_0 = 3.16\pm0.04~{\rm pc},
\end{equation}
Given that we have seen that narrower phase-space density distributions correspond to higher $r_0$ values, it is not surprising that the King radius we extract from the data is so much larger than the accepted value.

Recall that the point of this exercise is {\it not} to devise the most accurate single measurement of mass and radius of M4  -- for that purpose there are far better approaches to the {\it Gaia} data. The most accurate mass measurement would leverage the fact that M4 is still gravitationally bound (though phase-space density measurements may still be useful in such cases). Rather, we are interested in a mass measurement that will work {\it regardless} of whether a system which was once gravitationally bound is still bound at the present time. Other than the assumption of spherical symmetry, nothing of our approach has used the fact that the cluster is still gravitationally bound to derive the mass; as such there is the possibility that this measurement technique can be applied to other systems which have been tidally disrupted, though the precise choice of cuts would have to be re-tuned for a system with different kinematics. In that regard, finding a mass within 20\% of the accepted value is a surprising level of accuracy, and indicates that the significant effects of realistic measurement errors can be overcome. This could be made more accurate for the M4 globular cluster by better modeling of the effects of clustering on the measurement errors. However, this would be an unnecessary distraction from the main focus of this paper, which is developing this mass-measurement technique for disrupted objects (which in any event do no suffer from clustering effects to the degree that a still-bound globular cluster does).

\section{Entropy from Orbital Errors \label{sec:orbits}}

In the previous section, we have demonstrated that the phase-space density distribution of a gravitationally-bound cluster of stars can be used to accurately reconstruct the mass (and to a lesser extent, other profile parameters) despite measurement errors. Though measurement errors inject entropy into the system, a core of the phase-space overdensity remains, and can be recovered while removing foreground stars.

We now turn to a separate, but similar, problem which appears for gravitationally-disrupted systems. Here, measuring the phase-space density and volume is further complicated due to the dispersal of the tracer stars over vast distances with widely varying characteristic length scales. As a result of this dispersal, using position-velocity phase-space variables to measure the density tends to produce results which are incorrect by orders of magnitude. While this obstacle might be overcome by specialized kernels and metrics, or by dividing streams into small subsections with manageable length-scales (at the cost of increasing the impact of edge-effects), an existing method to accurately measure the phase-space volume is to use a set of canonical phase space coordinates other than the position and velocity basis. Instead, we turn to the integrals of the isolating integral actions $\vec{J}$ and the action angles $\vec{\theta}$ of a star's orbit in the Galaxy, which have been shown to be a more convenient set of coordinates for stellar streams \cite{Tremaine:1998nk,Helmi:2008eq}. 

However, these coordinates require a potential within which to calculate the orbit, and an incorrect potential will act as an entropy source -- driving the stars away from their kinematically-cold phase-space configuration \cite{2012ApJ...760....2P}. However, as we will demonstrate, the inevitable phase-space volume increase allows us to estimate the correct potential by minimizing the volume. This is motivated by Refs.~\cite{2008MNRAS.386L..47B,2012ApJ...760....2P,Sanderson:2014apa,2014ApJ...794....4P}, which note that the correct Galactic potential could be inferred by minimizing the dispersion of tidally-stripped substructure, and similar works measuring the Galactic potential through the kinematics of cold streams \cite{Johnston:1998bd,2013MNRAS.436.2386L,2013ApJ...778L..12P,2014MNRAS.437.2230M,2017A&A...601A..37B}.

For a quasi-periodic integrable orbits in non-chaotic systems, the action integrals can be defined over the orbit \cite{1987gady.book.....B,1978mmcm.book.....A}
\begin{equation}
J_i = \frac{1}{2\pi} \oint_{\gamma_i} \vec{v}\cdot d\vec{x}.
\end{equation}
The integral is performed over a path $\gamma_i$ over the torus defined by the orbit of the star. For loop orbits (orbits which themselves rotate around the center of the potential), each action can be interpreted as the oscillation of the orbit's angular momentum in some direction: choosing the three components of $\vec{J}$ as the radial action $J_r$, the $z$ angular momentum $L_z$, and $L-L_z$. The angles $\vec{\theta}$ are the canonical conjugates of the $\vec{J}$.

These coordinates have the very useful property that $d\vec{J}/dt = 0$, and $d\vec{\theta}/dt = $\,constant. However, calculating the action integrals and phase angles for an arbitrary potential is not possible. In this paper, we use the \textsc{Galpy} code \cite{2015ApJS..216...29B}, which implements the algorithms of Refs.~\cite{2012MNRAS.426.1324B,2012MNRAS.426..128S,Sanders:2014fma}. The action integrals for a loop orbit in an arbitrary potential is numerically estimated by solving a constrained system at each time step for the actual orbit.

For a cluster of stars being tidally stripped, these variables remain compact throughout the evolution: the action integrals $\vec{J}$ are constant over the orbit, and the angles $\vec{\theta}$ do not disperse throughout the bounded intervals $[0,2\pi]$. Coarse-graining will eventually become an issue, but this is not likely to be a concern for identifiable streams and tidal debris. 

\begin{figure}[t]
\includegraphics[width=0.4\columnwidth]{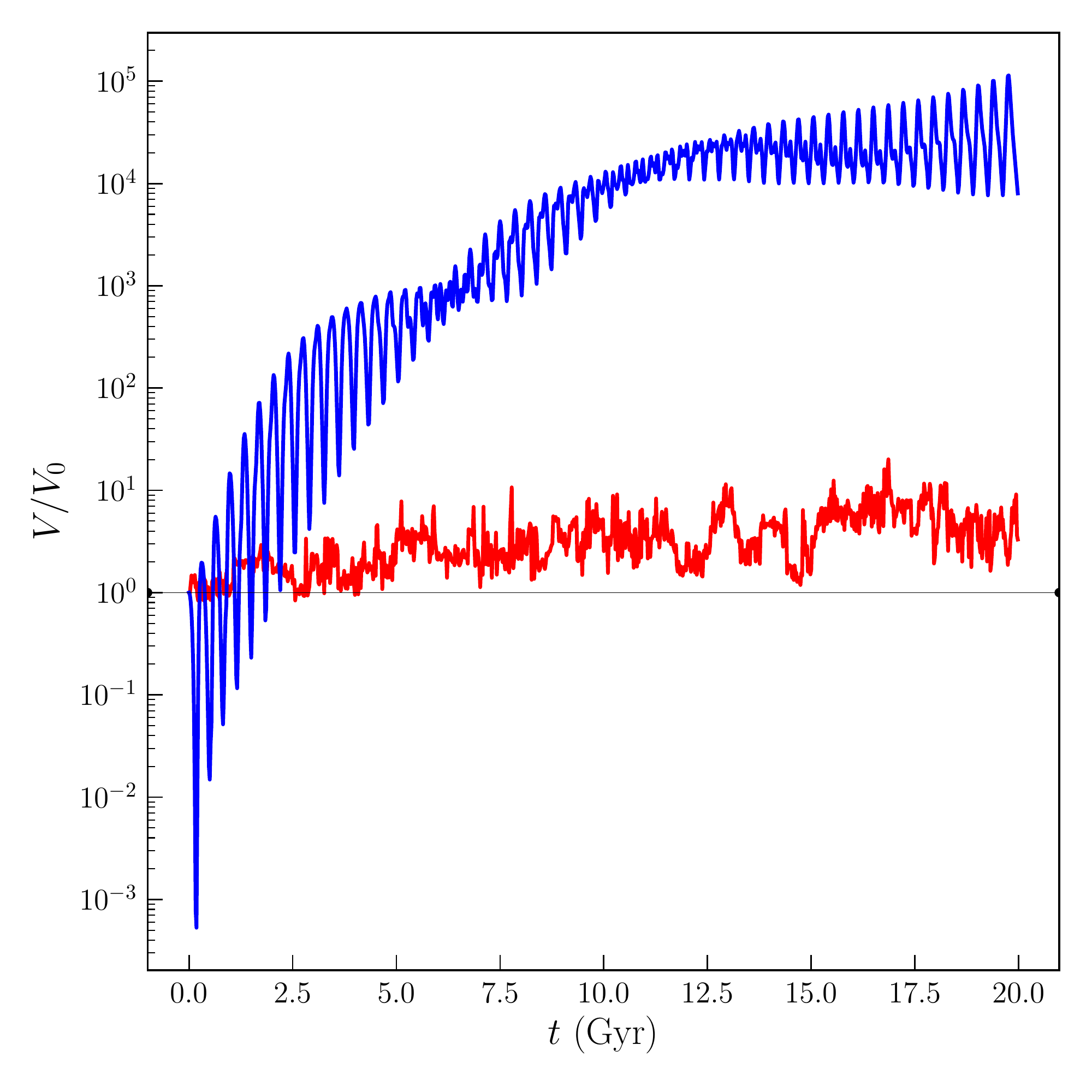}
\caption{Measured phase-space volume (divided by the initial volume) of 1000 stars evolving in an isochrone potential with a period of $\sim 500$~Myr, using position-velocity variables (blue) and action-integrals and angles (red). Stars are evolved using \textsc{Galpy} \cite{2015ApJS..216...29B}, and the volume is calculated as the sum over the inverse number densities of each star, as calculated using \textsc{EnBiD}: $V = \sum_i n_i^{-1}$. \label{fig:action_evolution}}
\end{figure}

The need for a different set of coordinates beyond position-velocity is demonstrated in Figure~\ref{fig:action_evolution}, where we show the evolution of the measured phase-space volume of a simulated cluster of 1000 stars as they are tidally stripped in an isochrone potential over 20 gigayears. We use an isochrone potential in this example, rather than an NFW profile because isochrone potentials allow analytic calculations of actions and phase angles, reducing the impact of numerical errors in the orbital evolution.\footnote{Note that this is not a full $N$-body simulation: the cluster lacks internal gravity and evolves only due to the potential of the galaxy. Nevertheless, Liouville's theorem still holds for the stars, and phase space volume should be conserved.} The orbital period for this system is approximately 500 Myr, so the system evolves through $\sim 40$ full orbits. Using position-velocity variables, the volume quickly increases by many orders of magnitude. In the action-angle space the system's volume can be accurately measured for many dynamical times (though numerical errors in our phase space measurement do result in an ${\cal O}(1)$ change in the phase space volume).\\

Clearly, action-angle variables are of great use to measuring the phase space volume of tidally-stripped systems. However, there remains a major hurdle which must be overcome: these variables are defined for an {\it orbit}, not for a particular value of position and velocity. To calculate an orbit from the position and proper motion measured by {\it Gaia} or similar survey requires an additional piece of information: the potential of the Milky Way Galaxy itself. 

Though many models of the total Galactic potential exist \cite[e.g.,][]{McMillan:2017}, these potentials are not fully accurate reproductions of the real Galaxy. As a result, the orbits derived for a stellar stream will generically be inaccurate, as will the action-integrals and angles. {\it A priori}, this mismodeling of the potential could either increase or decrease the phase space volume as measured in action-angle space, and would render this set of phase space coordinates useless for accurate measurements of the phase space volume.

\begin{figure}[t]
\includegraphics[width=0.45\columnwidth]{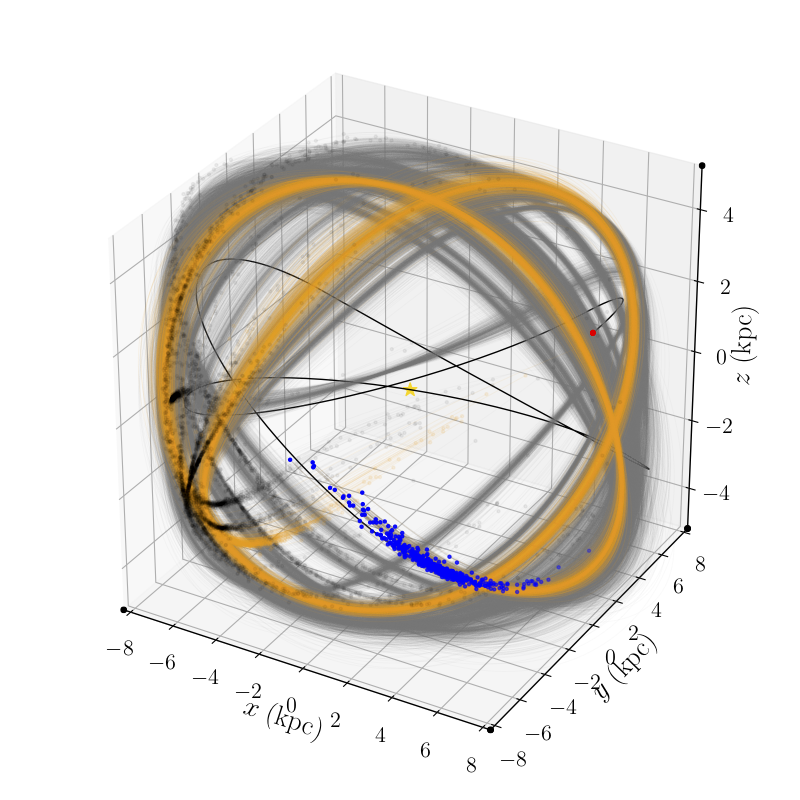}\includegraphics[width=0.45\columnwidth]{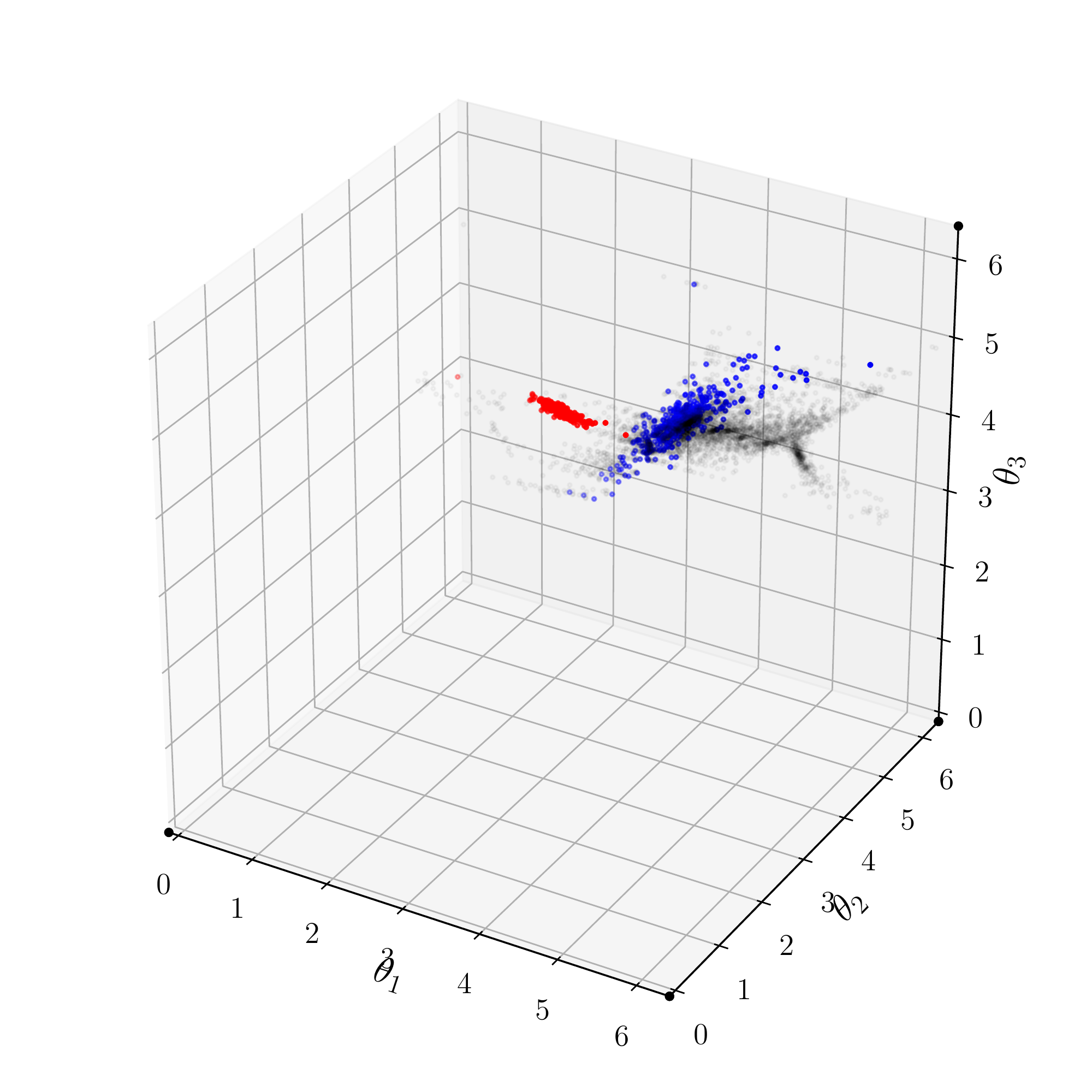}
\caption{Left: Evolution in position-space of a globular cluster being tidally-stripped in a Milky Way potential, starting as a King profile at $t=0$ (red dots). The solid black line shows the orbit from $t=0$ to 500 Myr (final locations shown as blue dots). The true orbit (in the correct Galactic potential) over the next 500 Myr is shown in orange, with ten orbits using incorrect potentials shown as grey lines ending in black dots. The center of the galaxy is denoted with a yellow star. Right: Evolution of this cluster in action-angle space from $t=0$ (red) to 500 Myr (blue). Ten representative sets of actions calculated in incorrect potentials are shown as black dots. \label{fig:action_example}}
\end{figure}

However, just as with mismeasurement errors (which only increase the phase-space volume), errors in the potential through which stars are orbited act as entropy sources for the phase-space density, and thus only {\it increase} the phase-space volume and associated entropy. As pointed out in Refs.~\cite{2008MNRAS.386L..47B,2012ApJ...760....2P,Sanderson:2014apa}, the potential in which a cluster of stars evolved to form cold streams can be recovered from within a larger set of incorrect potentials by looking at the compactness of the cluster in the action space. The correct potential is the one that results in orbits and actions that is the most compact in action space. In the language of this paper, the correct potential is the one that minimizes the total phase-space volume or entropy of the star cluster.

\begin{figure}[t]
\includegraphics[width=0.4\columnwidth]{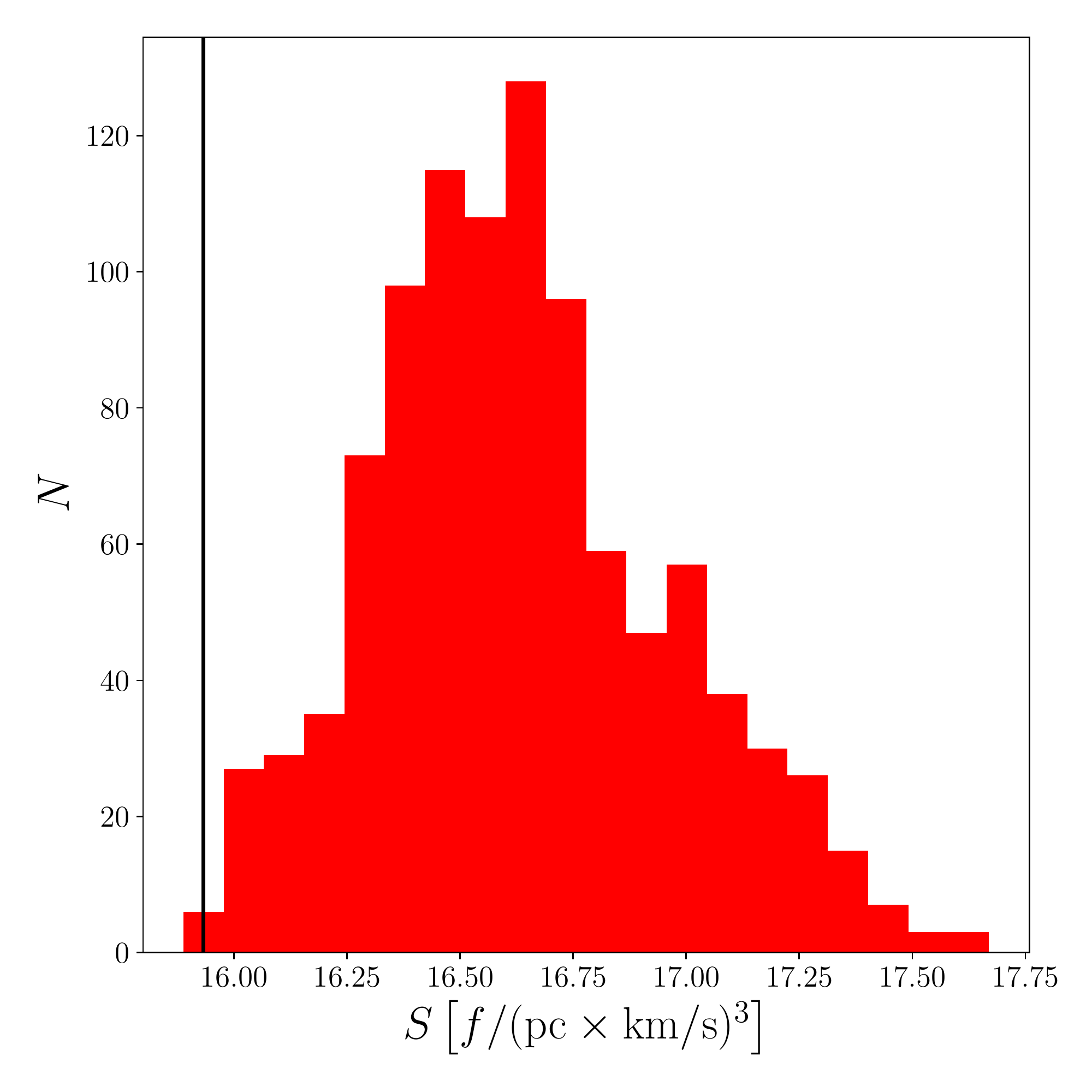}\includegraphics[width=0.4\columnwidth]{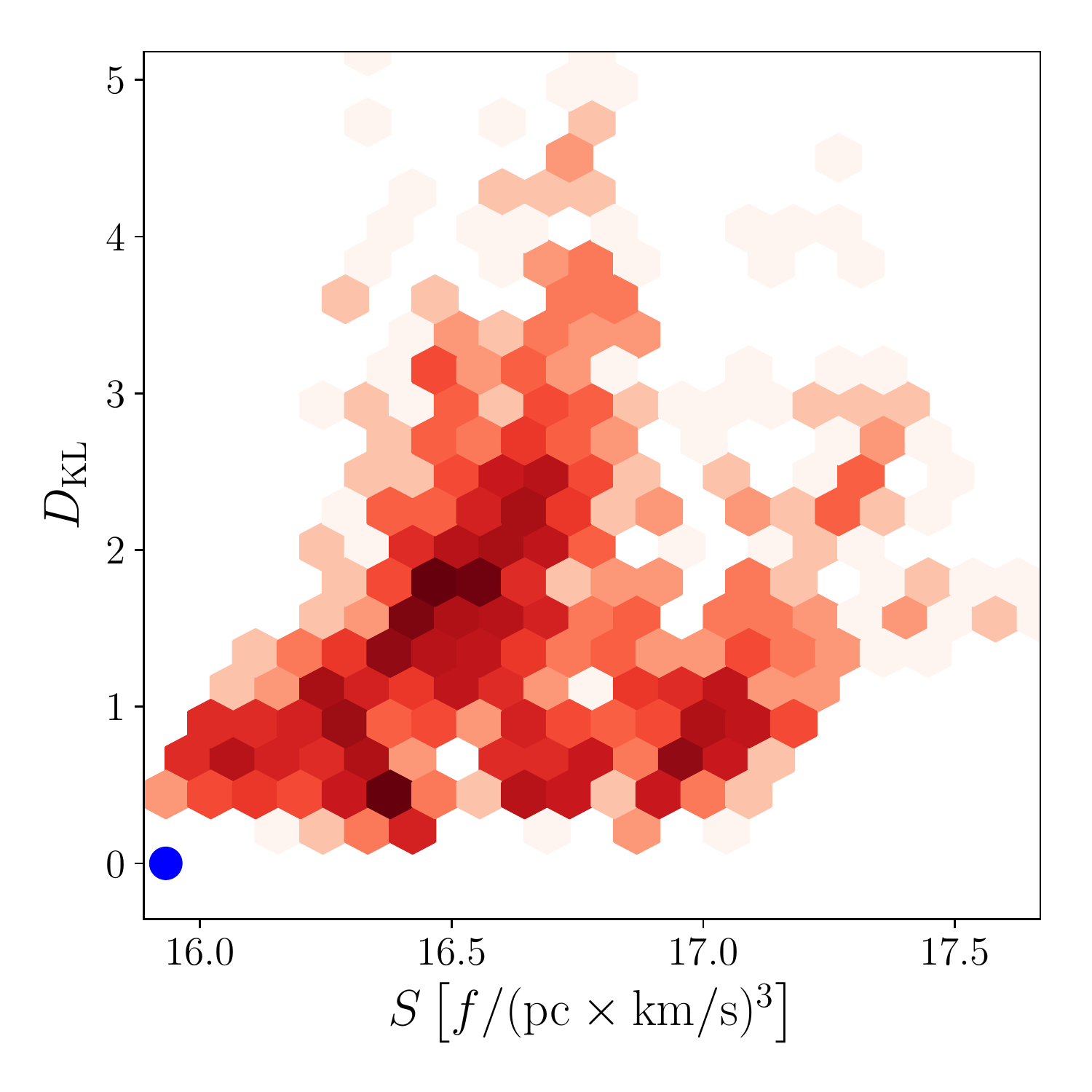}
\caption{Left: Histogram of the entropy of the phase-space density of 1000 stars calculated using action-angles derived from orbits in perturbations of a Milky Way-like potential (a combination of NFW, Miyamoto-Nagai, and power-law spherical potential with a cutoff). The true entropy, corresponding to orbits in the baseline potential which resulted in the initial conditions used in the perturbed orbits, is shown as a black vertical line. Right: Density plot of entropy versus KL test statistic between the true potential and the perturbed potential. Entropy corresponding to true potential shown in blue. \label{fig:orbitminimization}}
\end{figure}

Starting with the cluster of simulated stars (drawn from the King profile using the M4 best-fit parameters), we evolve them through 500~Myr in a particular Milky Way-like potential, creating a disrupted stream. We then create 1000 randomized potentials which are similar, but not identical, to the original potential. To do so, we treat the Milky Way potential as a sum of a NFW, Miyamoto-Nagai, and power-law spherical potential, and randomly perturb the shape parameters and relative normalization away from the original potential. For each potential, we find the orbits for each star, using the positions and velocities of the stars at 500 Myr after initial cluster disruption as the orbit's initial conditions (see the left panel of Figure~\ref{fig:action_example}). The phase-space density distribution is then calculated in the action-angle space using \textsc{EnBid}.\footnote{The \textsc{EnBid} kernel parameters in action-angle space are tuned so that the density distribution agrees with that measured in physical coordinates before cluster disruption.} Examples of the distribution in angle space are shown in the right panel of Figure~\ref{fig:action_example}. From this density distribution, we calculate the entropy using the definition Eq.~\eqref{eq:entropy_def}, replacing the volume integral over a discretized sum over the tracer stars. The histogram of the measured entropy for each choice of potential is shown in the left panel of Figure~\ref{fig:orbitminimization}, while the right panel shows the entropy versus the Kullback-Leiblier (KL) test statistic comparing the mass density distribution of the true potential and the perturbed potential inside the orbit of the cluster:
\begin{equation}
D_{\rm KL}(\rho_{\rm pert.}|\rho_{\rm true}) = \int_0^{10~\rm kpc} 2\pi r \int_{-5~\rm kpc}^{5\rm kpc} dz \,\frac{\rho_{\rm true}(r,z)}{M_{\rm true}}\ln\left[ \rho_{\rm pert.}(r,z)/ \rho_{\rm true}(r,z) \right], \label{eq:kltest}
\end{equation}
where $M_{\rm true}$ is the mass enclosed inside the orbit of the cluster. The key observation here is that the true phase-space volume is the minimum of all the tested potentials, and that perturbations which have phase-space volumes closer to the true answer are more similar to the ``real'' potential as measured by the KL statistic (zero corresponding to perfect agreement).

\begin{figure}[th]
\includegraphics[width=0.5\columnwidth]{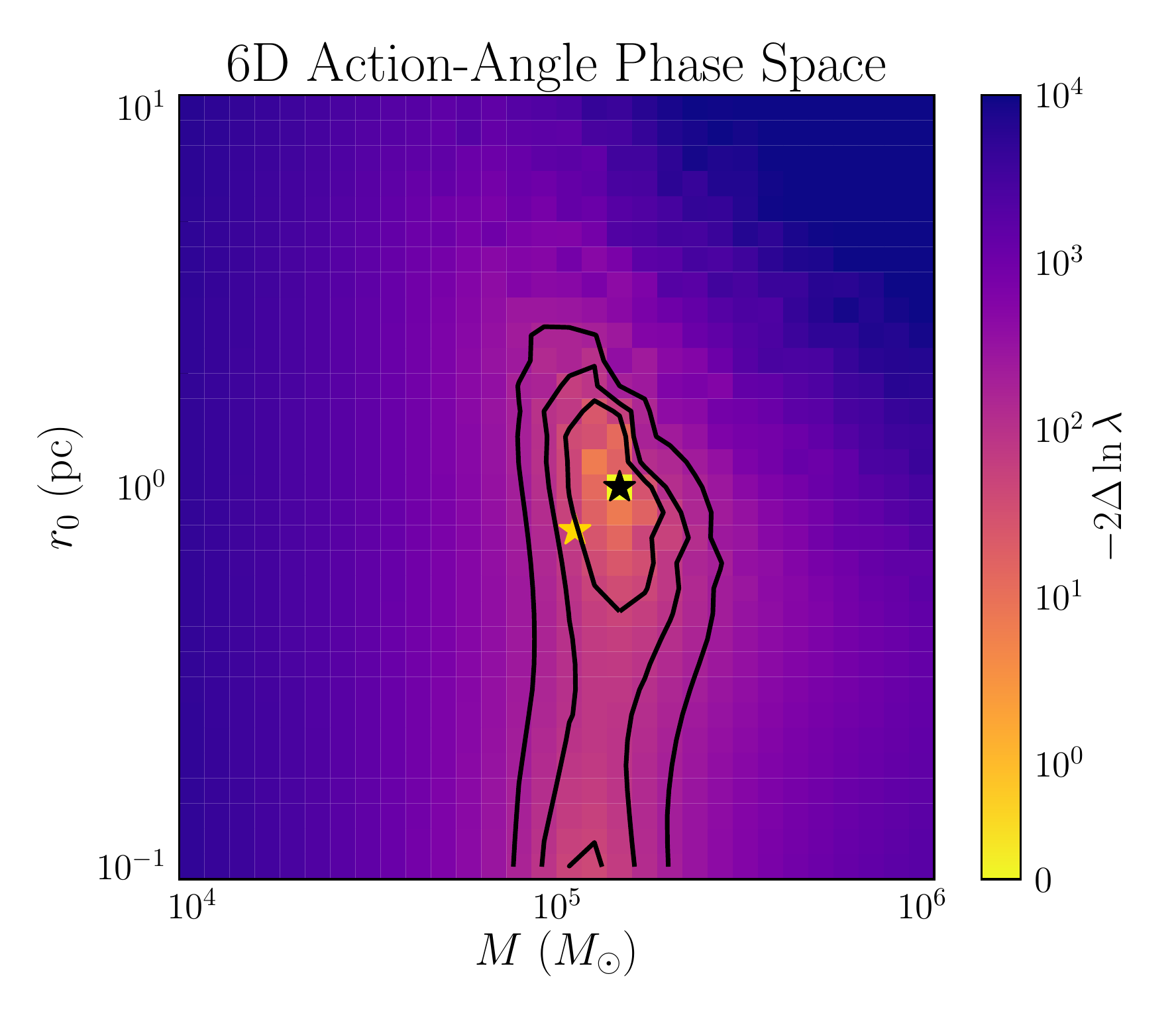}
\caption{Log-likelihood $-2\ln\lambda$ as a function of cluster mass $M$ and King radius $r_0$ (varying the concentration $c$ in order to keep $r_t = 34.9$~pc fixed) for a disrupted cluster after 500~Myr of evolution in Milky Way-like potential, using the phase-space density distribution reconstructed assuming a potential that minimizes the entropy. We use 1000 tracer stars without additional errors on the probability density values. The true value of the mass and King radius is shown with a gold star, the location of the minimum of $-2\ln\lambda$ is indicated with a black star. Contour lines are shown for reference, but do not represent $1\sigma$, $2\sigma$, {\it etc} ellipses.  \label{fig:disruptedreconstruction}}
\end{figure}

Of our 1000 iterations of the potential, we choose the one that minimizes the entropy of the cluster, and then -- as in the previous Section -- use the distribution of phase-space density to estimate the mass and original King radius of the cluster before disruption (for calculational simplicity, we assume that the tidal radius is known so that we do not have to scan over the concentration $c$). The resulting $-2\ln\lambda(\vec{\xi})$ over $\{M,r_0\}$ is shown in Figure~\ref{fig:disruptedreconstruction}, with a best-fit mass and $r_0$ of
\begin{equation}
M = (1.49 \pm 0.04)\times 10^5\,M_\odot,~r_0 = 1.00\pm0.05~{\rm pc}.
\end{equation}
We should expect that measurement errors will add additional systematic shifts away from the true values. However, it is notable that we can recover enough information from the action-angle space to make a relatively accurate measurement of the system's original mass and density profile, even in the simplified scenario considered here. Having demonstrated proof-of-principle, additional work will be required to extract six-dimensional phase space information with reasonable errors from the known streams, and we will continue this in a future paper.

\section{Conclusions \label{sec:conclusions}}

Liouville's theorem provides a powerful handle by which we may access the initial state of a previously-bound system after it has been tidally disrupted. This is especially interesting in light of {\it Gaia} data, which is uncovering streams and other substructure within our Galaxy. However, a number of hurdles present themselves which will confound accurate measurements of the phase space volume in realistic situations. In this paper, we consider two of these effects: measurement errors and orbital errors, and demonstrate how they can be overcome and the original distribution of phase-space density recovered. This can then be mapped back to the progenitor star cluster's mass and other structural parameters of the density profile.

Both measurement and orbital errors are effectively entropy injections into the system, which act to disperse the stars in six-dimensional phase space. That the entropy of the phase-space distribution always increases due to these errors provides us the key to recovering an accurate approximation of the original distribution. For measurement uncertainties, by considering stars with small measurement errors, the core overdensity remains intact, and the original distribution can be recovered by considering only the increasing density gradients for each star. With this, we have demonstrated using {\it Gaia} DR2 data that we can measure the mass and King radius of the nearest globular cluster with reasonable accuracy. As a side application, this technique may also be useful in determining cluster membership. 

Having shown that phase-space density can be used to determine the structure of stellar clusters which are still gravitationally bound, we turn to the much harder problem of disrupted systems. Given the difficulty of reconstructing accurate phase-space density using physical positions and velocities, alternative canonical coordinate systems such as action integrals and angles are better suited to the problem. However, these require an orbit for each star, which in turn requires knowledge of the Galactic potential in which the stars are orbiting.

As has been previously noted in the literature, the phase-space volume of a low-entropy collection of stars is minimized when the orbits are evolved in the true potential. We can therefore imagine recovering the true phase-space density by varying over possible potentials which approximate the Milky Way, minimizing the entropy associated with the density distribution. This makes action-angle space a viable choice of coordinates for measuring phase-space density and volume. Additional work will be needed to include effects of the bar and spiral arms on the orbits, as has already been noted in the context of Galactic streams \cite[e.g.,][]{Hattori:2016, PW:2016, Pearson:2017}, as well as including the measurement errors on the stream stars from {\it Gaia}.

In addition, recovering the potential through minimization of a phase-space entropy may allow probes of the inner shape of the Milky Way's (or other galaxies') dark matter profile. This possibility will be considered in future work.

Clearly a great deal of work remains before Liouville's theorem can be used to calculate the original properties of tidally-disrupted clusters of stars. For example, we have not addressed the combination of measurement and orbital errors in this paper, or a full six-dimensional measurement of stream stars in {\it Gaia} data. However, this paper is a proof-of-principle that these entropy increases are not insurmountable problems, and with additional work, the substructure of the Galaxy being revealed by {\it Gaia} can be mapped back to the original objects, with important implications for the study of galaxy formation and dark matter particle physics.

\acknowledgements

MRB is supported by the DOE grant DE-SC0017811. We thank Alyson Brooks, Kathryn Johnston, Lina Necib, Robyn Sanderson, David Spergel, and the attendees of the CCA Stars meeting for helpful conversation and advice. The calculations used in this paper were made using the \textsc{NumPy} \cite{numpy}, \textsc{SciPy} \cite{Jones:2001gn}, \textsc{AstroPy} \cite{astropy:2013, astropy:2018}, \textsc{Gala} \cite{Gala}, and \textsc{Galpy} \cite{2015ApJS..216...29B} packages for \textsc{Python3}.

\bibliography{phasespace}
\end{document}